\documentclass[12pt]{article}
\usepackage{epsf}
\usepackage{a4}
\usepackage{amssymb}
\usepackage{cite}

\def\one{
\setlength{\unitlength}{0.45cm}
\begin{picture}(0.55,0.5)
\put(0,0){\line(1,0){0.4}}
\put(0,.4){\line(1,0){0.4}}
\multiput(0,0)(.4,0){2}{\line(0,1){.4}}
\end{picture}}
\def\twohor{
\setlength{\unitlength}{0.45cm}
\begin{picture}(1.1,0.5)
\put(0,0){\line(1,0){0.8}}
\put(0,.4){\line(1,0){0.8}}
\multiput(0,0)(.4,0){3}{\line(0,1){.4}}
\end{picture}}
\def\twover{
\setlength{\unitlength}{0.45cm}
\begin{picture}(0.55,0.5)
\put(0,0){\line(1,0){0.4}}
\put(0,.4){\line(1,0){0.4}}
\put(0,-.4){\line(1,0){0.4}}
\multiput(0,0)(.4,0){2}{\line(0,0){.4}}
\multiput(0,0)(.4,0){2}{\line(0,-1){.4}}
\end{picture}}
\def\threehor{
\setlength{\unitlength}{0.45cm}
\begin{picture}(1.5,0.5)
\put(0,0){\line(1,0){1.2}}
\put(0,.4){\line(1,0){1.2}}
\multiput(0,0)(.4,0){4}{\line(0,1){.4}}
\end{picture}}
\def\threever{
\setlength{\unitlength}{0.45cm}
\begin{picture}(0.6,0.5)
\put(0,0){\line(1,0){0.4}}
\put(0,.4){\line(1,0){0.4}}
\put(0,-.4){\line(1,0){0.4}}
\put(0,-.8){\line(1,0){0.4}}
\multiput(0,0)(.4,0){2}{\line(0,0){.4}}
\multiput(0,0)(.4,0){2}{\line(0,-1){.4}}
\multiput(0,0)(.4,0){2}{\line(0,-2){.8}}
\end{picture}}
\def\mixed{
\setlength{\unitlength}{0.45cm}
\begin{picture}(1,0.5)
\put(0,0){\line(1,0){0.8}}
\put(0,.4){\line(1,0){0.8}}
\put(0,-.4){\line(1,0){0.4}}
\multiput(0,0)(.4,0){3}{\line(0,1){.4}}
\multiput(0,0)(.4,0){2}{\line(0,-1){.4}}
\end{picture}}
\def\denom{
\frac{1}{q^{{1\over 2}}-q^{-{1\over 2}}}}
\def\denomneg{
-\frac{1}{q^{{1\over 2}}-q^{-{1\over 2}}}}

\catcode`@=11 \@addtoreset{equation}{section} \catcode`@=12

\begin{document}
\begin{titlepage}
\noindent
{\tt TIFR/TH/00-53} \hfill
{\tt hep-th/0009188}\\
\hfill September, 2000 
\vfill
\begin{center}
{\Large \bf On Link Invariants and Topological String Amplitudes}\\[1cm] 
P. Ramadevi\footnote{Email: rama@phy.iitb.ernet.in}\\
{\em Physics Department, \\Indian Institute of Technology Bombay,\\
Mumbai 400 076, India\\[10pt]}
Tapobrata Sarkar\footnote{Email: tapo@theory.tifr.res.in}\\
{\em Department of Theoretical Physics, \\ Tata Institute of 
Fundamental Research, \\ Homi Bhabha Road, Mumbai 400 005, India}\\ 
\end{center}
\vfill
\begin{abstract}
We explicitly show that the new polynomial invariants for knots, 
upto nine crossings, agree with the Ooguri-Vafa conjecture 
relating Chern-Simons gauge theory to topological string
theory on the resolution of the conifold.
\end{abstract}
\vfill
\end{titlepage}

\section{Introduction}

Recently, there has been a lot of progress in understanding $SU(N)$
Chern-Simons theory on $S^3$ and its related invariants, the knot 
invariants \cite{wit,knotsgen}, from a topological string theory 
point of view \cite {gv1,gv2,ov,lm}. In a striking conjecture, 
Ooguri and Vafa \cite{ov} have proposed a formulation of knot 
invariants on $S^3$ from considerations of topological string theory 
amplitudes on the resolved conifold, based on earlier work by Gopakumar 
and Vafa\cite{gv1,gv2} that is similar in spirit to the celebrated 
AdS/CFT correspondence.  This conjecture, which was tested for the 
simplest knot, called the unknot (a circle in $S^3$) in \cite{ov} and 
for the trefoil knot (toral knot of type (2,3)~) by Labastida and Marino 
\cite{lm}, provides an extremely interesting ``physical'' interpretation 
for the coefficients in the  polynomial invariants. The purpose of this 
paper is to critically examine this conjecture for a wide class of knots 
using a simpler method of directly computing invariants for any knot or 
link carrying arbitrary representations of $SU(N)$ presented in Refs. 
\cite {gkr,thesi}.

In \cite{ov}, the structure of the knot invariants was deduced from
topological string amplitudes in the presence of D-2 branes 
ending on D-4 branes, and the result was expressed in terms of 
certain integers representing the number of D-2 branes. This constitutes
a physical interpretation of the integer coefficients appearing in two-variable
HOMFLY polynomials for knots. In \cite{lm}, 
this prediction was tested on the Chern-Simons field theory
side for toral knots (with explicit results for the right handed
trefoil knot) using the formulation of the toral knot operators 
introduced in \cite{lab1}. In particular, the
authors in \cite{lm} developed a group theory argument 
in relating the vacuum expectation values (vevs) of operators
on the topological string theory side to invariants of knots 
carrying arbitrary $SU(N)$ representation in Chern-Simons theory.
Their explicit calculations for the 
right handed trefoil knot agreed with Ooguri-Vafa conjecture  
on the polynomials appearing in the vev of such operators\cite{ov}. 
The computation becomes very tedious for 
knots with higher number of crossings within the toral knot
operator formalism.

We will use a simpler method \cite {gkr} for directly obtaining the 
invariants for any knot and link (includes torus
knots) carrying arbitrary $SU(N)$ representations. 
Then, incorporating the group theoretic results presented 
in Ref. \cite{lm}, we will compute the new polynomial invariants 
for a wide class of knots 
confirming Ooguri-Vafa conjecture. We can
evaluate multi-component link invariants within Chern-Simons
field theoretic framework but their relations
to the observables on the topological
closed string theory needs to be determined to check 
the conjecture. 

%

The organisation of the paper is as follows. In the first part of 
section 2, we briefly recapitulate the salient features of the conjecture 
in \cite{ov}. Next, we review the procedure of \cite{lm} in computing
new polynomial invariants from invariants of knots carrying 
arbitrary $SU(N)$ representation. 
In section 3, we briefly present the method of directly calculating 
knot invariants for arbitrary representations of $SU(N)$ \cite {gkr}. 
In section 4, we present our results on 
knot and link invariants for arbitrary representations, upto 
9 crossings. In section 5, we
conclude with discussions and comments on some open problems. 

\section{Topological String Theory Amplitudes and its Relation to 
Chern-Simons Gauge theory}

In this section, we begin by briefly recapitulating the essential details
of the argument of \cite{ov,gv1} which proposes an equivalence
between $SU(N)$ Chern-Simons gauge theory on $S^3$ and closed A-twisted 
topological string theory on the $S^2$ resolved conifold geometry.
For a large $N$ gauge theory, in 'tHooft's double line notation, denoting
by $g$ the genus of the (triangulated) Reimann surface formed by the 
double lines, and by $h$ the number of faces in the triangulation of 
this surface the free energy can be written as the summation
\begin{equation}
F=\sum_{g,h}C_{g,h}N^{2-2g}{\tilde\lambda}^{h-2+2g}
\nonumber
\end{equation}
where $\tilde{\lambda}$ is the 'tHooft coupling related 
to the perturbative gauge
coupling constant $\kappa$ by ${\tilde\lambda}=\kappa N$. For the case of
Chern-Simons theory on $S^3$, the coefficients $C_{g,h}$ were shown to 
be equal to the partition function of the A-twisted open topological 
string theory with $g$ handles and $h$ boundaries with the target space
as the cotangent space of $S^3$ \cite{wittencs}. In the context of 
D-branes, this implies that the Chern-Simons theory is equivalent to
the open topological string theory obtained by wrapping $N$ D-branes 
on the base $S^3$ of the cotangent space. On the other hand, we can  
consider the cotangent space $T^*S^3$ to be the deformed conifold that
can, via a conifold transition, be resolved into the total space of
the line bundle ${\cal O}(-1)+{\cal O}(-1)$ over $S^2$. In \cite{gv2},
it was conjectured that the closed topological string theory on this
resolved conifold geometry is dual to the $SU(N)$ Chern-Simons theory
on $S^3$ with the K\"ahler parameter $t$ of the blown up $S^2$ being 
related by $t=\frac{2\pi iN}{k+N}$ to the coupling constant $k$ of
the Chern-Simons theory. Elegant computations were done in \cite{gv2},
(following earlier work done by these authors in \cite{gv1}) in order
to check this conjecture at the level of the free energy (the vacuum
amplitude $Z$ of the gauge theory being related to the free energy as
$Z=e^{-F}$), and their results strongly supported the said conjecture.

In \cite{ov}, this conjecture was further extended to the level of 
observables of the Chern-Simons theory, which are the gauge invariant 
and metric independent Wilson loop operators $W_R(C)$ associated with a 
knot $C$, (which is an embedding of a circle $S^1$ on a three manifold),
and carrying an irreducible representation $R$ of the gauge group, in 
our case the gauge group being $SU(N)$. These operators are defined in 
the usual way as trace of the holonomy $U(C)$:
\begin{eqnarray}
W_R(C)&=&{\mbox{Tr}}_R {\mbox {U(C)}} \\
{\rm where}~ U(C)&=&
{\mbox{P}}{\mbox{exp}}
\left[\oint_CA_{\mu}dx^{\mu}\right]
\end{eqnarray}
with ${\mbox{Tr}}_R$ being the trace over the representation $R$ and
P denotes path ordering.  
The vacuum expectation value (vev) of the Wilson loop operator
$W_R (C)$ gives the knot invariant $V_R[C]$ for knot $C$ embedded in $S^3$:
\begin{equation}
V_R[C]~=~\langle W_R(C) \rangle ~=~{ \int [{\cal D}A_{\mu}] W_R(C) \exp (iS)
\over \int [{\cal D}A_{\mu}] \exp (iS) }~. \label {2five}
\end{equation}
Here $S$ denotes the Chern-Simons field theory action:
\begin{equation}
S = ({k\over 4\pi}) \int _{S^3} Tr \left(A\wedge dA + {2\over 3} A\wedge
A\wedge A \right ) ~, \label {2one}
\end{equation}
where $A$ is a one form valued in the Lie algebra of
$SU(N)$ gauge group and $k$ is the coupling constant
which takes integer values.
Besides the Wilson loop associated with a single curve $C$ (knot), 
we could consider product of Wilson loops for a  
link $L~=~\prod_i^s C_i$ made up of component knots
$C_1, C_2, \ldots C_s$ carrying
representation $R_1, R_2, \ldots R_s$ as observables. The invariant
for such a multi-component link  
is given by
\begin{equation}
V_{R_1, R_2, \ldots R_s} [L]~=~ \langle W_{R_1, R_2, \ldots,R_s} (L)
\rangle ~=~{ \int [{\cal D}A_{\mu}] \prod_i
W_{R_i}(C_i) \exp (iS)~\over \int [{\cal D}A_{\mu}]
\exp (iS) }. \label {2six}
\end{equation}
For $R_i$'s chosen to be defining representation of $SU(N)$,
the invariant gives two-variable HOMFLY polynomial \cite {homf}
(upto unknot normalisation) which reduces to 
the celebrated Jones' polynomials for $SU(2)$ gauge group.
We will present the essential ingredients for directly
evaluating these knot and link invariants in the next section.

The observables in the topological string theory on a deformed confiold, 
are constructed using the following
basic idea \cite{ov}: Define a Lagrangian submanifold (a 
3-cycle on which the symplectic form vanishes) corresponding to a given 
knot $C$ in $S^3$ such that $C$ represents the intersection of the
3-cycle with $S^3$. Wrapping $M$ D-branes on the 3-cycle 
now gives a Chern-Simons
theory on the 3-cycle in addition to the $SU(N)$ Chern-Simons theory
on $S^3$, and quantizing the action for the complex scalar field 
on $C$ gives rise to the effective operator
\begin{equation}
Z(U,V)={\mbox{exp}}\left[\sum_{n=1}^{\infty}\frac{1}{n}{\mbox{Tr}}U^n
{\mbox{Tr}}V^{-n}\right] \label {oper}
\end{equation} 
where $U$ and $V$ are the holonomies
transforming under $SU(N)$ and $SU(M)$ groups respectively 
and the trace is over the defining representation. 
Clearly, these operators involve traces of powers of holonomy 
besides the Wilson loops and products of Wilson loops
which has not been considered within Chern-Simons field
theoretic framework. We will see later in this section that the vev of 
these new operators can be related to Chern-Simons invariants for 
knots carrying arbitrary $SU(N)$ representation.

The form of the vev of $Z(U,V)$ was conjectured by invoking
duality of this theory with closed topological string theory on the 
resolved conifold. This was done using a method analogous to the 
Schwinger computation of 
the free energy of a particle in a background field. 
It is known \cite {agnt} that in type IIA compactification on Calabi-Yau, the 
topological string amplitudes at genus $g$ compute corrections to 
the low energy theory of the form $R_+^2F_+^{2g-2}$ (with $R_+$ and
$F_+$ being the self dual field strengths of the curvature and the 
graviphoton respectively). These can be computed via a one-loop
calculation where the relevant D-brane states, that are D-2 branes 
wrapped on cycles bound to D-0 branes, are integrated out.  
The Lagrangian submanifold on which D-branes are wrapped in the 
Chern-Simons calculation is identified on the resolved conifold geometry, 
and the Schwinger computation is done with $M$ D-4 branes wrapping
the Lagrangian 3-cycle and extending along an ${\bf R}^2$ in non-compact
space time, on which the D-2 branes might end. The $U(1)$ gauge field
on the non-compact ${\bf R}^2$ gives rise to a magnetic two form field
on the D-4 brane, and each D-brane state that enters in the 
(two dimensional) Schwinger calculation is characterised by its magnetic 
charge $R$ (with $R$ transforming in some representation of 
the $U(M)$ gauge theory on the D-4 brane), bulk D-2 brane charge $Q$ and 
spin $s$. Denoting the number of such states by $N_{R,Q,s}$, 
the closed topological string amplitude is \cite{ov}
\begin{eqnarray}
\langle Z(U,V)\rangle&=&{\mbox{exp}}\sum_{n=1}^{\infty}\sum_R
f_R(q^n,\lambda^n){\mbox{Tr}}_R\frac{V^n}{n}
\label{lmeq}\\
{\rm where}~f_R(q,\lambda)&=&\sum_{s,Q}
\frac{N_{Q,R,s}}{q^{1/2}-q^{-1/2}}\lambda
^Qq^s \label{lmaeq}~,
\end{eqnarray}
where \footnote{Our $q$ is the variable $t$ defined in \cite{lm}
and equal to $e^{i\lambda}$ of \cite{ov}.}
$q={\mbox{exp}}\left[\frac{2\pi i}{k+N}\right]$; $\lambda=q^N$. 

In order to obtain the above conjectured form (\ref {lmeq}) 
from Chern-Simons theory, we need to device methods
of computing vevs of trace of powers of holonomy.
An elegant group theoretic approach \cite {lm} 
proves useful in arriving at the form (\ref {lmeq})
and also obtaining the new polynomial invariant $f_R(q, \lambda)$ 
in terms of invariants of knots carrying arbitrary $SU(N)$ representation
in Chern-Simons theory. We will see in section 4 that these
$f_R(q, \lambda)$ for a wide class of knots indeed 
obey eqn. (\ref {lmaeq}).

The basic idea involved in the group theoretic method is to use 
the Frobenius formula (see, for eg. \cite{fultonharris})  
to compute product of traces of arbitrary powers of the 
holonomy. Introducing a class function $\Gamma_{\vec k} (U)$, 
this formula reads as
\begin{equation}
\Gamma_{\vec k}(U)=\prod_{j=1}^{\infty}\left({\mbox{Tr}}U^j\right)^{k_j}  
=\sum_{Y \in S_{\ell}} \chi_Y({\bf C( \vec k)}){\mbox{Tr}}_{R(Y)}U
\label{lmbasis}
\end{equation}
where ${\bf C(\vec k)}$ denotes the conjugacy class determined by the
sequence $(k_1,k_2, \cdots)$ (i.e there are $k_1$ 1-cycles, 
$k_2$ 2-cycles etc) in the permutation group $S_{\ell}$    
(${\ell}=\sum_j jk_j$), $Y \in S_{\ell}$ is the Young Tableau with
$n= \ell$ number of boxes, $R(Y)$ is the $SU(N)$ representation 
corresponding to the Young Tableau $Y$.  $\chi_Y ({\bf C(\vec k)})$
gives the character of the conjugacy class ${\bf C(\vec k)}$
in the representation $Y$ of the permutation group $S_{\ell}$.
Rewriting the exponent in the  eqn. (\ref {oper}) 
and ${\rm Tr}_R V^n$ in the eqn. (\ref {lmeq})  
in terms of the class functions, the following
equality can be obtained \cite {lm}:
\begin{eqnarray}
{\rm log } \langle Z (U,V) \rangle &=&
\sum_{\vec k} {\vert C(\vec k) \vert \over
{\ell}!}  G_{\vec k}^{(c)} (U) \Gamma_{\vec k} (V)  \label {compa} \\
~&=& \sum_{\vec k} {\vert C(\vec k) \vert \over {\ell}!}  
\sum_{n \vert \vec k} n^{\vert k \vert -1} \sum_R \chi_R ({\bf
C(\vec k_{1 \over n}})) f_R (q^n, \lambda^n) \Gamma_{\vec k} (V) \nonumber
\end{eqnarray} 
where $n \vert \vec k$ means  `` $n$ divides $\vec k$ ''
and $(\vec k_{1 \over n})_i = \vec k_{ni}$. 
For example,
$\vec k_{1 \over n} = (k_n, k_{2n}, \cdots)$
for the vector $\vec k = (0,0, \cdots, k_n, 0, \cdots, k_{2n}, \cdots).$
and the ``connected'' vev $G_{{\bf k}}^{(c)}$
is defined through the relation  
\begin{equation}
{\mbox{log}}\left[1+\sum_{{\bf \vec k}}\frac{\vert C({\bf \vec k}) \vert}{l!}
G_{{\bf k}} \prod_jx^{k_j}\right]=\sum_{{\bf \vec k}}\frac{\vert 
C({\bf \vec k}) \vert}{l!}
G_{{\bf k}}^{(c)}\prod_jx^{k_j} \label {recurs}  
\end{equation}
where $G_{{\bf k}}=\langle\Gamma_{{\bf k}}\rangle$, 
$|C({\bf k})| =\ell ! / (\prod_j ~k_j !~ j ^{k_j})$ is
the number of elements in the conjugacy class ${\bf C(\vec k)}$.
From eqn. (\ref {compa}), we deduce
\begin{equation}
G_{{\bf k}}^{(c)}(U)=\sum_{n|{\bf k}}n^{|{\bf k}|-1}\sum_R
\chi_R({\bf C(\vec k_{1 \over n})}) f_R(q^n,\lambda^n)
\label{masterlm}
\end{equation}
Using eqns. (\ref {lmbasis}, \ref {recurs}, \ref {masterlm}),  
$f_R$ can be expressed in 
terms of the vev of ${\mbox{Tr}}_R(U)$ and certain lower order terms.
We refer the reader for more details to \cite{lm}, but for the moment,
we list the formal expressions for $f_R$ for

\noindent 
\setlength{\unitlength}{0.7cm}
$R=
\begin{picture}(0.55,0.5)
\put(0,0){\line(1,0){0.4}}
\put(0,.4){\line(1,0){0.4}}
\multiput(0,0)(.4,0){2}{\line(0,1){.4}}
\end{picture},
\setlength{\unitlength}{0.7cm}
\begin{picture}(1.1,0.5)
\put(0,0){\line(1,0){0.8}}
\put(0,.4){\line(1,0){0.8}}
\multiput(0,0)(.4,0){3}{\line(0,1){.4}}
\end{picture},
\setlength{\unitlength}{0.7cm}
\begin{picture}(0.55,0.5)
\put(0,0){\line(1,0){0.4}}
\put(0,.4){\line(1,0){0.4}}
\put(0,-.4){\line(1,0){0.4}}
\multiput(0,0)(.4,0){2}{\line(0,0){.4}}
\multiput(0,0)(.4,0){2}{\line(0,-1){.4}}
\end{picture},
\setlength{\unitlength}{0.7cm}
\begin{picture}(1.5,0.5)
\put(0,0){\line(1,0){1.2}}
\put(0,.4){\line(1,0){1.2}}
\multiput(0,0)(.4,0){4}{\line(0,1){.4}}
\end{picture},
\setlength{\unitlength}{0.7cm}
\begin{picture}(0.6,0.5)
\put(0,0){\line(1,0){0.4}}
\put(0,.4){\line(1,0){0.4}}
\put(0,-.4){\line(1,0){0.4}}
\put(0,-.8){\line(1,0){0.4}}
\multiput(0,0)(.4,0){2}{\line(0,0){.4}}
\multiput(0,0)(.4,0){2}{\line(0,-1){.4}}
\multiput(0,0)(.4,0){2}{\line(0,-2){.8}}
\end{picture},
{\mbox{and}}~
\setlength{\unitlength}{0.7cm}
\begin{picture}(1,0.5)
\put(0,0){\line(1,0){0.8}}
\put(0,.4){\line(1,0){0.8}}
\put(0,-.4){\line(1,0){0.4}}
\multiput(0,0)(.4,0){3}{\line(0,1){.4}}
\multiput(0,0)(.4,0){2}{\line(0,-1){.4}}
\put(1.1,0){.}
\end{picture}$
\vskip0.5cm
\noindent
For the single box representation
of $SU(N)$, the invariant
$f_{
\setlength{\unitlength}{0.45cm}
\begin{picture}(0.55,0.5)
\put(0,0){\line(1,0){0.4}}
\put(0,.4){\line(1,0){0.4}}
\multiput(0,0)(.4,0){2}{\line(0,1){.4}}
\end{picture}}$ is simply given by the expectation value of $U$ in
the single box representation, i.e 
\begin{equation}
f_{\one}=\langle{\mbox{Tr}}_{\one}U\rangle
\label{onebox}
\end{equation}
The r.h.s of the above equation is the unnormalised HOMFLY polynomial,
as we will come to in the next section. Similarly, for the 
horizontal and vertical double box representations of $SU(N)$, we get 
from eq. (\ref{masterlm}),
\begin{eqnarray}
f_{\twohor}(q,\lambda)&=&\langle{\mbox{Tr}}_{\twohor}U\rangle-{1\over 2}
f_{\one}(q,\lambda)^2-{1\over 2}f_{\one}(q^2,\lambda^2)
\nonumber\\
f_{\twover}(q,\lambda)&=&\langle{\mbox{Tr}}_{\twohor}U\rangle-{1\over 2}
f_{\one}(q,\lambda)^2+{1\over 2}f_{\one}(q^2,\lambda^2)
\label{twoboxes}
\end{eqnarray}
and similarly, for representations of $SU(N)$ with three boxes in
the Young Tableau, we get the equations
\begin{eqnarray}
f_{\threehor}(q,\lambda)&=&\langle{\mbox{Tr}}_{\threehor}U\rangle
-f_{\one}(q,\lambda)f_{\twohor}(q,\lambda)-{1\over 6}f_{\one}(q,\lambda)
^3 \nonumber \\
&-&{1\over 2}f_{\one}(q,\lambda)f_{\one}(q^2,\lambda^2)
-{1\over 3}f_{\one}(q^3,\lambda^3)
\nonumber\\
f_{\mixed}(q,\lambda)&=&\langle{\mbox{Tr}}_{\mixed}\rangle
-f_{\one}(q,\lambda)f_{\twover}(q,\lambda)-f_{\one}(q,\lambda)f_{
\twohor}(q,\lambda)-{1\over 3}f_{\one}(q,\lambda)^3\nonumber\\
&+&{1\over 3}f_{\one}(q^3,\lambda^3)
\nonumber\\
f_{\threever}(q,\lambda)&=&\langle{\mbox{Tr}}_{\threever}U\rangle
-f_{\one}(q,\lambda)f_{\twover}(q,\lambda)-{1\over 6}f_{\one}
(q,\lambda)^3+{1\over 2}f_{\one}(q,\lambda)f_{\one}(q^2,\lambda^2)
\nonumber\\
&-&{1\over 3}f_{\one}(q^3,\lambda^3)
\label{threeboxes}
\end{eqnarray}
The above equations can be used to find the values of $f$ in various
representations in a recursive manner. For example, the quantity
$f_{\one}=\langle{\mbox{Tr}}_{\one}U\rangle$ is, as we have mentioned,
 the unnormalised HOMFLY polynomial. Using this and the value of 
$\langle{\mbox{Tr}}_{~ 
\setlength{\unitlength}{0.45cm}
\begin{picture}(1.1,0.5)
\put(0,0){\line(1,0){0.8}}
\put(0,.4){\line(1,0){0.8}}
\multiput(0,0)(.4,0){3}{\line(0,1){.4}}
\end{picture}}~U\rangle(q,\lambda)$ (the method of calculation 
will be presented in the next section) we can evaluate
$f_{\setlength{\unitlength}{0.45cm}
\begin{picture}(1.1,0.5)
\put(0,0){\line(1,0){0.8}}
\put(0,.4){\line(1,0){0.8}}
\multiput(0,0)(.4,0){3}{\line(0,1){.4}}
\end{picture}}(q,\lambda)$ from eq. (\ref{twoboxes}). We can then use
this value of 
$f_{\setlength{\unitlength}{0.45cm}
\begin{picture}(1.1,0.5)
\put(0,0){\line(1,0){0.8}}
\put(0,.4){\line(1,0){0.8}}
\multiput(0,0)(.4,0){3}{\line(0,1){.4}}
\end{picture}}(q,\lambda)$
in eq. (\ref{threeboxes}) to calculate $f_{\threehor}$ using the 
expression for $\langle{\mbox{Tr}}_{\threehor}U\rangle$.
In ref. \cite{ov}, the conjecture (\ref{lmeq}),(\ref{lmaeq}) was
verified for the simplest curve called unknot:
\begin{eqnarray}
\langle Z(U,V)\rangle&=&
{\mbox {exp}}\sum_{n=1}^{\infty} 
f_{
\setlength{\unitlength}{0.45cm}
\begin{picture}(0.55,0.5)
\put(0,0){\line(1,0){0.4}}
\put(0,.4){\line(1,0){0.4}}
\multiput(0,0)(.4,0){2}{\line(0,1){.4}}
\end{picture}} (q^n, \lambda^n) {{\rm tr} V^n \over n} \\
{\rm where}~~f_{ \setlength{\unitlength}{0.45cm}
\begin{picture}(0.55,0.5)
\put(0,0){\line(1,0){0.4}}
\put(0,.4){\line(1,0){0.4}}
\multiput(0,0)(.4,0){2}{\line(0,1){.4}}
\end{picture}} (q, \lambda)&=&(\lambda^{1\over 2}-
\lambda^{-\frac{1}{2}})\over (q^{1 \over 2}-q^{-\frac{1}{2}})
\end{eqnarray}
implying $f_R = 0$ for representations other than the 
defining representation. 

For a class of toral knots (explicit results for right-handed trefoil),
$f_R (q , \lambda)$ was shown to agree with the conjecture (\ref {lmaeq})
\cite {lm}.

In the next section, we will present our direct method of
computing ${\rm Tr}_R U$ for a wide class of knots and links
(which includes torus knots) carrying arbitrary 
representation of $SU(N)$ which will be used in section 4 to
evaluate $f_R (q, \lambda)$ and hence 
proving Ooguri-Vafa conjecture for such knots.

\def\myfigure#1#2{
\begin{figure}\baselineskip=14pt plus 2pt minus 1pt
\centerline{#1}\nobreak\smallskip\nobreak \caption{#2}
\end{figure}}
\def\myfigures#1#2{
\begin{figure}
\centerline{#1}\nobreak\smallskip\nobreak \caption{#2}
\end{figure}}
\def\eps{\epsilon}
\def\a{\alpha}
\def\l{\lambda}
\def\b{\beta}
\def\w{\omega}
\def\be{\begin{equation}}
\def\ee{\end{equation}}
\def\ket#1{|{#1}\rangle}  
\def\bra#1{\langle{#1}|}  
\def\norm#1#2{\langle{#1}|{#2}\rangle}   
\def\gov#1#2#3#4{\pmatrix{{#1}&{#2}\cr{#3}&{#4}}}
\def\govi#1#2#3#4{\left[\matrix{{#1}&{#2}\cr{#3}&{#4}}\right]}
\section{Generalised Knot and Link Invariants from 
$SU(N)$ Chern-Simons Theory}

In this section, we  present the methods of obtaining 
invariants of knots and links carrying arbitrary $SU(N)$ representations
in Chern-Simons theory \cite {gkr}. For representations other than
defining representations placed on knots/links, we will 
refer to the invariants as generalised knot/link invariants. 

Explicit computations are done for
some knots and links obtained from braids which
includes toral knots of type $(2,2m+1)$. These invariants will
be useful in obtaining the new polynomial invariants 
which are related to amplitudes in topological string theory.

In order to directly evaluate the knot and link invariants, 
we require the  following two ingredients:
\begin{itemize}
\item The functional integral over the space of matrix valued one forms $A$ 
is evaluated by exploiting the connection between the 
Chern-Simons theory in three dimensional space with boundary
and the $SU(N)_k$ Wess-Zumino conformal field theory on the 
two dimensional boundary \cite {wit}. 
\item Any knot/link can be represented as a closure or plat or capping
of braids \cite {bir}.
\end{itemize}

The computation of these invariants has been considered in 
detail in  Refs. \cite{gkr, thesi}.
In order to illustrate the technique of direct
evaluation of the invariant, 
we will  concentrate on knots and links $L_m$ obtained as
a closure of a two-strand braid with $m$-crossings.
$L_m$ represents toral knots of type (2,m) for $m$ odd and two-component
links for $m$ even. 
The details we present for this class of knots and links
are general enough for deriving invariants of knots and links obtained 
from any $n$-strand braid by closure or plat or capping of braids.

\myfigure{\epsfxsize=2in \epsfbox{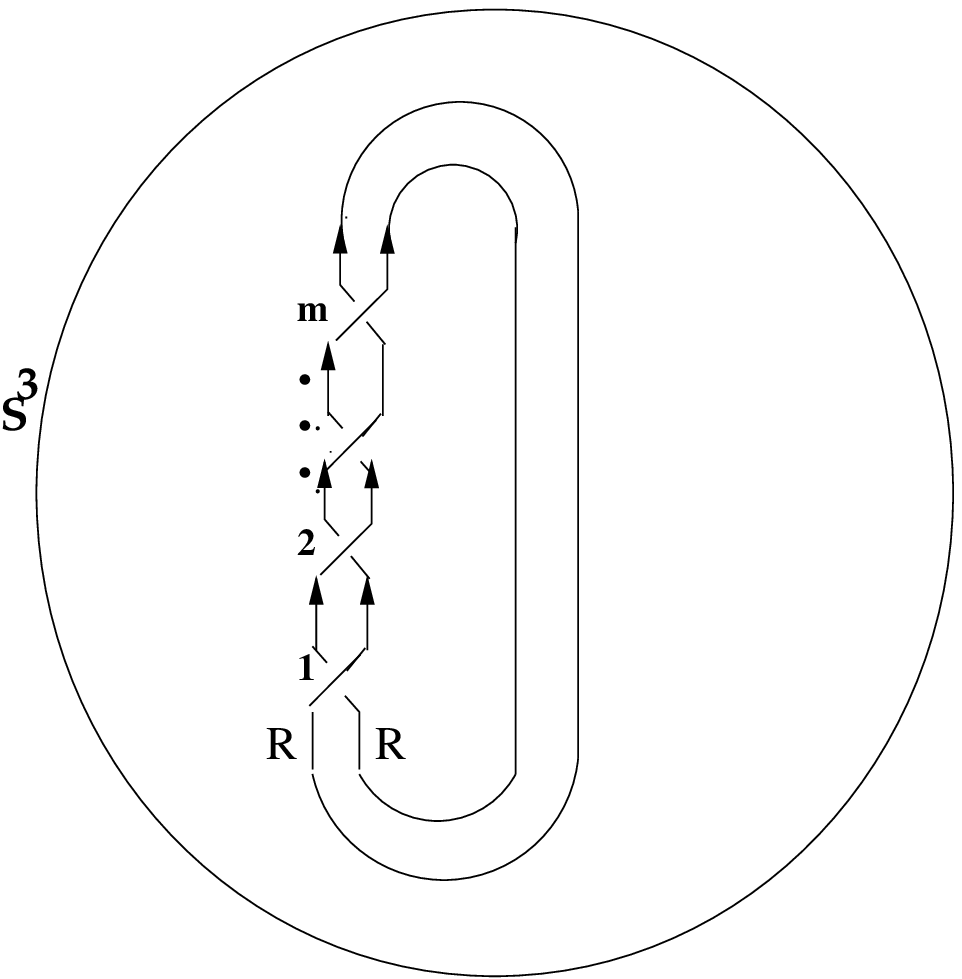}} {\sl $L_m$ obtained as closure
of Two-Strand Braid}
\myfigure{\epsfxsize=3in \epsfbox{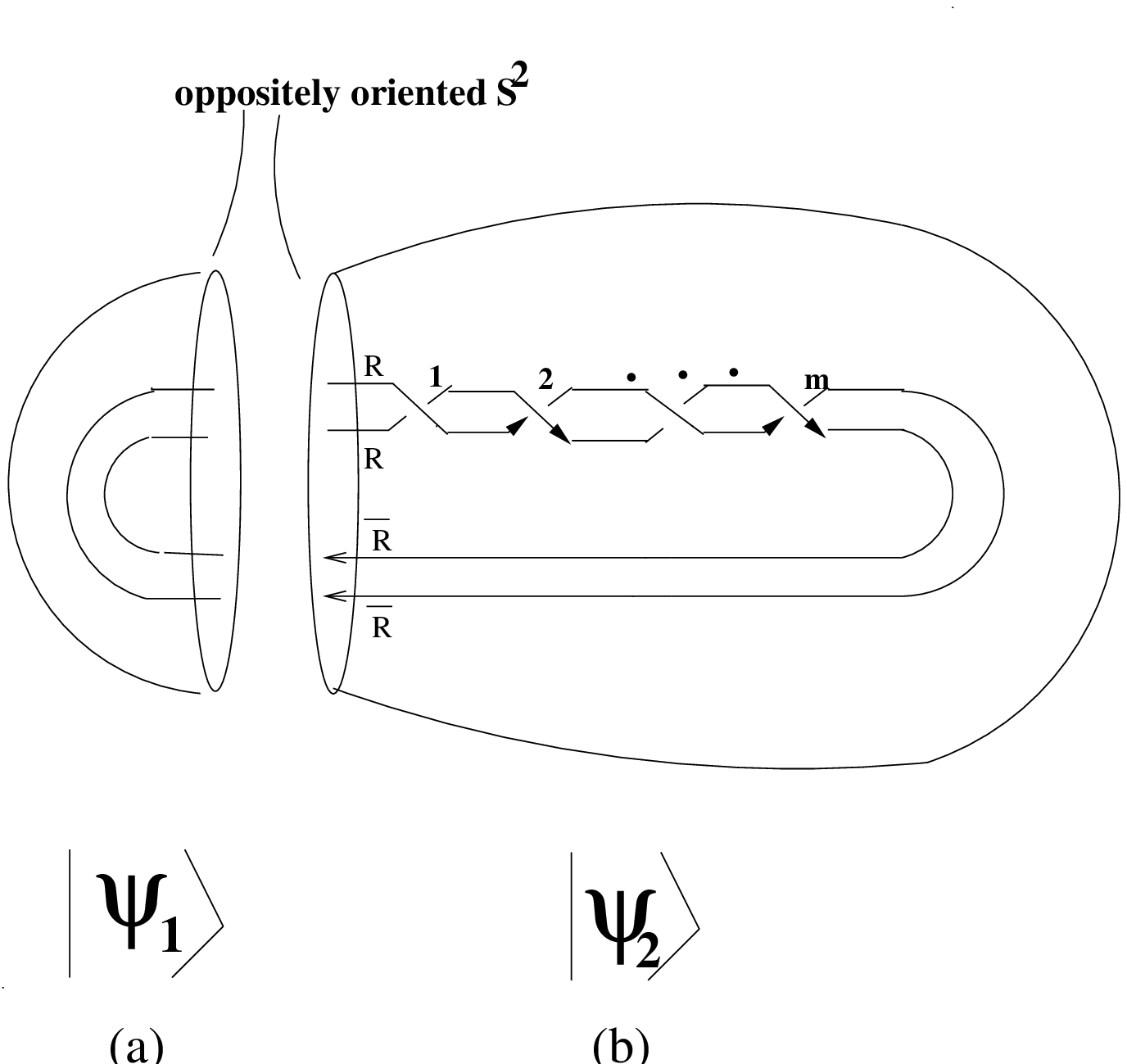}}{\sl Two three-balls with
oppositely oriented $S^2$ boundaries}
Consider the  three-manifold $S^3$ containing $L_m$ 
carrying representation $R$ as shown in Fig.~1.\footnote{
We could place two different representations $R_1, R_2$
in the two component knots of link $L_{2m}$.}
Let us slice the three-manifold into two three dimensional balls 
as shown in Fig.2(a) and (b). 
The two dimensional $S^2$ boundaries 
of the three-balls are oppositely oriented and have four 
points of intersections with the braid, which we refer to as
four punctures.
Now, exploiting the connection between 
Chern-Simons theory and Wess-Zumino conformal field theory,
the functional integrals of these three-balls  correspond to
states in the space of four point correlator conformal blocks
of the Wess-Zumino conformal field theory \cite{wit}. 
The dimensionality of this space is dependent on the representation 
of $SU(N)$ placed on the strands and the number of 
punctures on the boundary. 
These states can be written in a suitable basis. 
Two such choices of bases ($|\phi_s^{side} \rangle )$, and 
$(|\phi_t ^{cent} \rangle$) are pictorially depicted in 
Fig.3(a),(b). Here $s \in R \otimes R$ and $t \in R \otimes \bar R$ 
as allowed by the $SU(N)_k$ Wess-Zumino conformal field theory
fusion rules.
The basis $|\phi_s^{side} \rangle$ is chosen when the braiding is done
in the side two parallel strands. In other words, it is the 
eigen basis corresponding to the braiding 
generators $b_1$ and $b_3$:
\be
|\phi_s^{side}\rangle =
b_1 |\phi_s^{side}\rangle =
b_3 |\phi_s^{side}\rangle =
(\lambda_s^{(+)} (R,R)) 
|\phi_s^{side}\rangle
\ee
with the eigenvalues $\lambda_s ^{(+)}(R_1,R_2)$ for  
right-handed half-twists between two parallel strands 
carrying different representation $R_1, R_2$ being:
\be
\lambda_s^{(+)}(R_1,R_2) =
\epsilon_{R_1R_2}^s q^{C_{R_1} + C_{R_2}+ \vert C_{R_1} - C_{R_2} \vert 
/2- C_s/2}.
\label {pevalue}
\ee
where $\epsilon_{R_1R_2}^s = \pm 1$ and 
$C_{R_i}'s, C_s$ are the quadratic casimir in the 
representations $R_i's$ and $s$ respectively.
In terms of the highest weight $\Lambda_R$ for representation
$R$, the quadratic casimir is given by 
\begin{equation}
C_R = {1 \over 2} \left(\Lambda_R . (\Lambda_R + 2 \rho)\right)~, 
\end{equation}
where $\rho$ is the Weyl vector equal to the sum of the  
fundamental weights of $SU(N)$.
\myfigure{\epsfxsize=3in \epsfbox{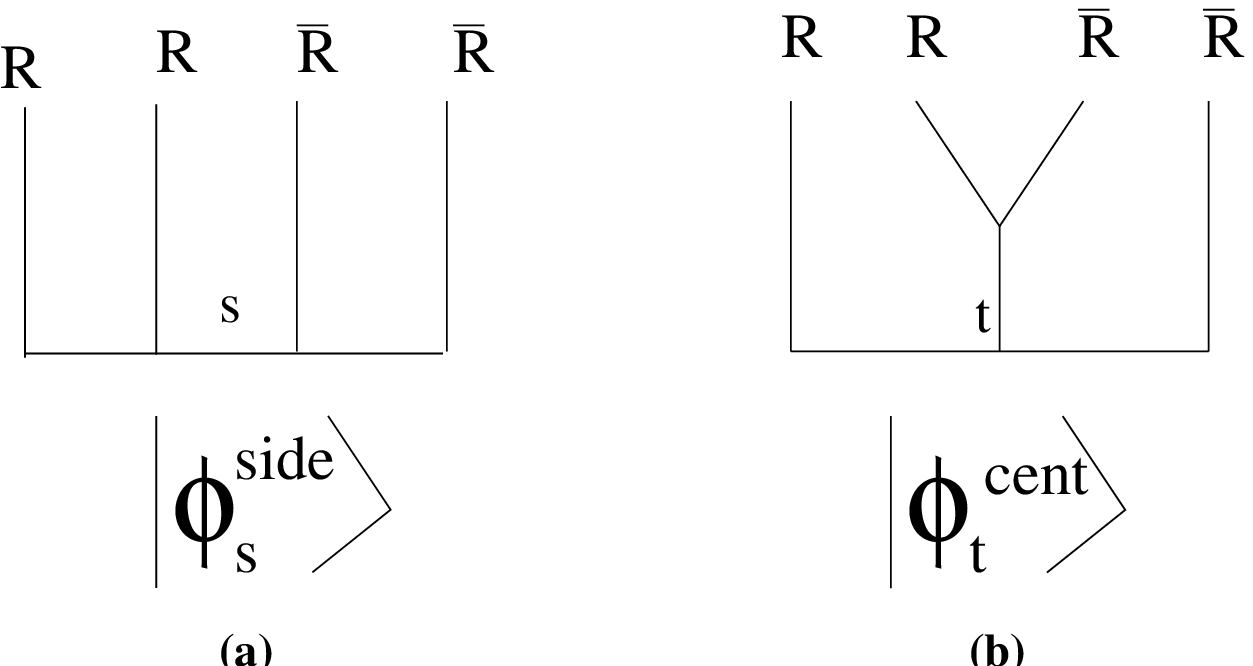}}{\sl Basis states on a
four-punctured $S^2$ boundary}

Similarly for braiding in the middle two anti-parallel strands $b_2$,
we choose the basis $|\phi_t^{cent} \rangle$:
\be
b_2 |\phi_t^{cent}\rangle = (\lambda_t^{(-)} (R,\bar R) 
|\phi_t^{cent}\rangle \label {aeval}
\ee
with the eigenvalues for right-handed half-twists in the 
anti-parallel strands being:
\be
\lambda_t^{(-)}(R_1, \bar R_2)=
{\hat \epsilon}_{R_1 \bar R_2}^t q^{-\vert C_{R_1} - C_{R_2} \vert/2+ C_t/2}
\ee
where ${\hat \epsilon}_{R_1\bar R_2}^t= \pm 1$.

These two bases are related by a duality matrix: 
\be
|\phi_s^{side} \rangle = a_{st}\left(\matrix{R &
R \cr \bar R & \bar R \cr}\right) |\phi_t^{cent} \rangle
\label {dual}
\ee
The matrix elements of the duality matrix are the $SU(N)$ quantum
Racah coefficients which are known for some special representations 
\cite {gkr}.

For example, the duality matrix for $R$ taken to be defining
representation of $SU(N)$) is a
$2 \times 2$ matrix:
\be
{1 \over [N]}\left( \matrix{\sqrt{[N][N-1] \over [2]} & \sqrt{[N][N+1] 
\over [2]} \cr
          \sqrt{[N][N+1] \over [2]} & -\sqrt{[N][N-1] \over [2]} \cr}\right)
\ee
where the number in square bracket refers to the quantum number defined as
\be
[n] = {q^{n \over 2} - q^{-{n \over 2}} \over 
q^{1 \over 2} - q^{-{1 \over 2}}} \label {quan}
\ee
with $q$ (also called deformation parameter in quantum
algebra $SU(N)_q$) related to the coupling constant $k$ as
$q = {\rm exp} ({2 i \pi \over k+N})$.
We will see that the invariants are polynomials in $q, \lambda= q^N$.

Now, let us determine the states corresponding to Figs.2(a) and (b).
Since the braiding is in the side two parallel strands, 
it is preferable to use $|\phi_s^{(side)} \rangle$ as basis states. 
Let $|\Psi_1 \rangle$ be the state corresponding to Fig. 2(a). 
Clearly, we can write the state for Fig. 2(b) as
\be
|\Psi_2 \rangle = b_1^m |\Psi_1 \rangle
\ee
This state should be in the dual space as its $S^2$ boundary
is oppositely oriented compared to the boundary in Fig. 2(a).
Then the link invariant is given by 
\be
V_R[L_m] ~=~\langle \Psi_1 |b_1^m |\Psi_1 \rangle \label {inv}
\ee
For determining the polynomial, we will have to express 
the states as linear combination of the basis 
states $|\phi_s^{side} \rangle$. The coefficients 
in the linear combination are chosen such that
$$\langle \Psi_1 | \Psi_1 \rangle =\left( V_R[U] \right) = 
({\rm dim}_q R)^2,$$ 
where the  quantum dimension for any irreducible representation 
$R$ with highest weight $\Lambda$ is given by
 \be
 dim_q R~=~\prod_{\a > 0} { [\a . (\rho + \Lambda)] \over [\a . \rho]}~,
\label {qdim}
 \ee
where $\a$'s are the positive roots and $\rho$ is equal to the sum of the
fundamental weights of the Lie group $SU(N)$. 
The square bracket refers to usual
definition of quantum number (\ref {quan}).

The above mentioned restrictions determine the state $|\Psi_1\rangle$
(see \cite {gkr}) as: 
\be
|\Psi_1\rangle 
 = \sum_{s \in R \otimes R} \sqrt{{\rm dim}_q s} |\phi_s^{side~} \rangle
\ee
Substituting it in eqn. (\ref {inv}) and using the braiding eigenvalue 
(\ref {pevalue}),
we obtain
\begin{equation}
V_R[L_m]~=~ \sum_{s \in R \otimes R} {\rm dim}_q R_s
 (\lambda_{s}^{(+)}(R,R))^m~. \label {knot}
\end{equation}
For a two-component link with different representations, the multi-coloured
link invariant will be \cite {gkr}
\begin{equation}
V_{R_1, R_2}[L_{2m}]~=~ \sum_{s \in R_1 \otimes R_2} {\rm dim}_q R_s
 (\lambda_{s}^{(+)}(R_1,R_2))^{2m}~. \label {link}
\end{equation}
\setlength{\unitlength}{.5cm} 
For $R=\begin{picture}(.7,.7)
\put(0,0){\line(1,0){.3}}
\put(0,.3){\line(1,0){.3}}
\multiput(0,0)(.3,0){2}{\line(0,1){.3}}
\end{picture}~$
denoting $SU(N)$  defining representation,
we get the HOMFLY polynomial $P_{L_m} [l , m]$ \cite {homf,lick}  
(upto unknot normalisation) 
\begin{equation}
V_{\begin{picture}(.4,.7)
\put(0,0){\line(1,0){.3}}
\put(0,.3){\line(1,0){.3}}
\multiput(0,0)(.3,0){2}{\line(0,1){.3}}
\end{picture}}[L_m] = 
{(\lambda^{1\over 2} - \lambda^{-1 \over 2}) \over 
(q^{1 \over 2} - q^{-1\over 2})} P_{L_m}[l = i\lambda, 
m = i (q^{1/2}- q^{-1/2})]~.
\end{equation}

So far, we have elaborated for a class of knots and
links obtained from closure of two-strand braids
with $m$ crossings and directly obtained the invariant 
(\ref{knot}, \ref{link}).
The essential ingredients presented for this class
is sufficient to derive the invariant for any knot/link obtained
from $n$-strand braids. 

Let us now determine explicitly the irreducible representations 
and braiding values for some $SU(N)$ representations. 

For symmetric representation placed on component knots
\setlength{\unitlength}{.7cm}
$R_n =
\begin{picture}(2,.7)
\put(0,0){\line(1,0){1.8}}
\put(0,.3){\line(1,0){1.8}}
\multiput(0,0)(.3,0){7}{\line(0,1){.3}}
\put(.7,.5){\vector(-1,0){.7}}
\put(1.1,.5){\vector(1,0){.7}}
\put(.8,.5){$n$}
\end{picture},$
the irreducible representations $\rho_{\ell} \in R_n \otimes R_n$ in 
$SU(N)_k$ Wess-Zumino conformal field theory in the large $k$ limit 
will be 
\setlength{\unitlength}{1cm}
$$\begin{picture}(10,0)
\put(0,0){\line(1,0){1.8}}
\put(0,.3){\line(1,0){1.8}}
\multiput(0,0)(.3,0){7}{\line(0,1){.3}}
\put(.7,.5){\vector(-1,0){.7}}
\put(1.1,.5){\vector(1,0){.7}}
\put(.8,.5){$n$}
\put(2,0){$\otimes$}
\put(.8,-.4){$R_n$}
\put(2.6,0){\line(1,0){1.8}}
\put(2.6,.3){\line(1,0){1.8}}
\multiput(2.6,0)(.3,0){7}{\line(0,1){.3}}
\put(3.3,.5){\vector(-1,0){.7}}
\put(3.7,.5){\vector(1,0){.7}}
\put(3.4,.5){$n$}
\put(3.4,-.4){$R_n$}
\put(4.8,0){$=$}
\put(5.4,0){$\oplus_{\ell=0}^n$}
\put(6.5,0){\line(1,0){3}}
\put(6.5,.3){\line(1,0){3}}
\put(6.5,-.3){\line(1,0){1.8}}
\multiput(6.5,0)(.3,0){11}{\line(0,1){.3}}
\multiput(6.5,0)(.3,0){7}{\line(0,-1){.3}}
\put(8,-.6){$\rho_{\ell}$}
\put(6.8,.5){\vector(-1,0){.3}}
\put(8,.5){\vector(1,0){.3}}
\put(6.9,.5){$n-\ell$}
\put(8.6,.5){\vector(-1,0){.3}}
\put(9.2,.5){\vector(1,0){.3}}
\put(8.8,.5){$2\ell$}
\end{picture}~.$$
The quantum dimension (\ref {qdim}) for $\rho_{\ell}$ and parallel braiding 
eigenvalues (\ref {pevalue}) for strands carrying symmetric 
representation $R_n$ can be rewritten in a neat form:
\begin{eqnarray}
{\rm dim}_q \rho_{\ell} &=& { ([N] [N+1] \ldots [N+n+\ell-1]) ([N-1][N] \ldots 
[N+n-\ell-2]) \over [2 \ell]!~ [n- \ell]! ~
([n+\ell+1] [n+\ell] \ldots [2 \ell +2])}~,\nonumber\\
\l^{(+)}_{\rho_{\ell}} (R_n, R_n) &=& (-)^{n-\ell}~q^{(N-1)n \over 2} 
q^{\frac{n(n+1)}{2} - \frac{\ell(\ell +1)}{2}} 
\end{eqnarray}
Similarly, for antisymmetric representations placed on the component knots,
the irreducible representations ${\hat \rho}_{\ell}$ will be
\vskip.5cm
$$\begin{picture}(10,1.5)
\put(0,0){\line(0,1){1.8}}
\put(.3,0){\line(0,1){1.8}}
\multiput(0,0)(0,.3){7}{\line(1,0){.3}}
\put(.5,.7){\vector(0,-1){.7}}
\put(.5,1.1){\vector(0,1){.7}}
\put(.5,.8){$n$}
\put(1,.8){$\otimes$}
\put(-.6,.8){${\hat R}_n$}
\put(2.3,0){\line(0,1){1.8}}
\put(2.6,0){\line(0,1){1.8}}
\multiput(2.3,0)(0,0.3){7}{\line(1,0){.3}}
\put(3.1,.7){\vector(0,-1){.7}}
\put(3.1,1.1){\vector(0,1){.7}}
\put(3.1,.8){$n$}
\put(1.7,.8){${\hat R}_n$}
\put(3.8,.6){$=$}
\put(4.2,.6){$\oplus_{\ell=0}^n$}
\put(5.8,2){\line(0,-1){3}}
\put(6.1,2){\line(0,-1){1.8}}
\put(5.5,-1){\line(0,1){3}}
\put(5.5,2){\line(2,0){.6}}
\put(5.5,1.7){\line(2,0){.6}}
\put(5.5,1.4){\line(2,0){.6}}
\put(5.5,1.1){\line(2,0){.6}}
\put(5.5,.8){\line(2,0){.6}}
\put(5.5,.5){\line(2,0){.6}}
\put(5.5,.2){\line(2,0){.6}}
\put(5.5,-.1){\line(1,0){.3}}
\put(5.5,-.4){\line(1,0){.3}}
\put(5.5,-.7){\line(1,0){.3}}
\put(5.5,-.7){\line(1,0){.3}}
\put(5.5,-1){\line(1,0){.3}}
\put(5.7,-1.5){${\hat \rho}_{\ell}$}
\put(6.2,1.2){\vector(0,1){.8}}
\put(6.2,1){$\ell$}
\put(6.2,.8){\vector(0,-1){.65}}
\put(6.2,-.6){\vector(0,-1){.5}}
\put(6.2,-.2){\vector(0,1){.5}}
\put(6.2,-.4){$2(n-\ell)$}
\end{picture}~.$$
\vskip1cm

\noindent
We can again rewrite the quantum dimension (\ref {qdim}) for 
${\hat \rho}_{\ell}$ and 
eigenvalues in parallel strands (\ref {pevalue}) carrying
${\hat R}_n$ representations as follows:
\begin{eqnarray}
{\rm dim}_q {\hat \rho}_{\ell}&=&{ ([N] [N-1] \ldots [N-2n+\ell+1]) 
([N+1][N] \ldots [N-\ell+2]) \over [2n-2\ell]!~ [\ell]!~
([2n-2 \ell+2] [2n-2 \ell+3] \ldots [2n-\ell+1])} \nonumber\\
\l^{+}_{{\hat {\rho}}_{\ell}} (\hat R_n, \hat R_n) &=&  (-)^{n-\ell}
q^{(N-1)n \over 2} ~q^{n- \ell (n+1)+ 
\frac{\ell(\ell +1)}{2}} ~.
\end{eqnarray}

Even though we have explicitly given the braiding eigenvalues 
for symmetric and antisymmetric representations, the general
procedure presented in this section can be used to obtain 
eigenvalues for strands carrying mixed symmetry representations
as well. For example $R= \begin{picture}(.7,.7)
\put(0,0){\line(2,0){.6}}
\put(0,.3){\line(2,0){.6}}
\multiput(0,0)(.3,0){3}{\line(0,2){.3}}
\put(0,-.3){\line(0,1){.3}}
\put(.3,-.3){\line(0,1){.3}}
\put(0,-.3){\line(1,0){.3}}~
\end{picture}$, 
the irreducible representations in the tensor-product will be
$$
\setlength{\unitlength}{.8cm}
\begin{picture}(10,-1)
\put(0,0){\line(2,0){.6}}
\put(0,.3){\line(2,0){.6}}
\multiput(0,0)(.3,0){3}{\line(0,2){.3}}
\put(0,-.3){\line(0,1){.3}}
\put(.3,-.3){\line(0,1){.3}}
\put(0,-.3){\line(1,0){.3}}
\put(.1,-.8){$R$}
\put (.65,0){$\otimes$}
\put(1.3,0){\line(2,0){.6}}
\put(1.3,.3){\line(2,0){.6}}
\multiput(1.3,0)(.3,0){3}{\line(0,2){.3}}
\put(1.3,-.3){\line(0,1){.3}}
\put(1.6,-.3){\line(0,1){.3}}
\put(1.3,-.3){\line(1,0){.3}}~
\put(1.4,-.8){$R$}
\put(2,0){$=$}
\put(2.5,0){\line(3,0){.9}}
\put(2.5,.3){\line(3,0){.9}}
\multiput(2.5,0)(.3,0){4}{\line(0,1){.3}}
\put(2.5,.3){\line(0,-4){1.2}}
\put(2.8,.3){\line(0,-4){1.2}}
\put(2.5,-.3){\line(1,0){.3}}
\put(2.5,-.6){\line(1,0){.3}}
\put(2.5,-.9){\line(1,0){.3}}
\put(2.9, -1){$\rho_1$}
\put(3.5,0){$\oplus$}
\put(3.9,0){\line(3,0){.9}}
\put(3.9,.3){\line(3,0){.9}}
\multiput(3.9,0)(.3,0){4}{\line(0,1){.3}}
\put(3.9,.3){\line(0,-3){.9}}
\put(4.2,.3){\line(0,-3){.9}}
\put(4.5,.3){\line(0,-2){.6}}
\put(3.9,-.3){\line(2,0){.6}}
\put(3.9,-.6){\line(1,0){.3}}
\put(4, -1){$\rho_2$}
\put(4.9,0){$\oplus$}
\put(5.3,0){\line(3,0){.9}}
\put(5.3,.3){\line(3,0){.9}}
\multiput(5.3,0)(.3,0){4}{\line(0,1){.3}}
\put(5.3,.3){\line(0,-3){.9}}
\put(5.6,.3){\line(0,-3){.9}}
\put(5.9,.3){\line(0,-2){.6}}
\put(5.3,-.3){\line(2,0){.6}}
\put(5.3,-.6){\line(1,0){.3}}
\put(5.4, -1){$\rho_3$}
\put(6.3,0){$\oplus$}
\put(6.7,0){\line(4,0){1.2}}
\put(6.7,.3){\line(4,0){1.2}}
\multiput(6.7,0)(.3,0){5}{\line(0,1){.3}}
\multiput(6.7,-.3)(.3,0){3}{\line(0,1){.3}}
\put(6.7,.3){\line(0,-2){.6}}
\put(7,.3){\line(0,-1){.3}}
\put(6.7,-.3){\line(2,0){.6}}
\put(6.8, -1){$\rho_4$}
\put(8,0){$\oplus$}
\put(8.4,0){\line(3,0){.9}}
\put(8.4,.3){\line(3,0){.9}}
\put(8.4,-.3){\line(3,0){.9}}
\put(8.4,.3){\line(0,-2){.6}}
\put(8.7,.3){\line(0,-2){.6}}
\put(9,.3){\line(0,-2){.6}}
\put(9.3,.3){\line(0,-2){.6}}
\put(8.6, -1){$\rho_5$}
\put(9.6,0){$\oplus$}
\put(10.2,0){\line(2,0){.6}}
\put(10.2,.3){\line(2,0){.6}}
\put(10.2,-.3){\line(2,0){.6}}
\put(10.2,-.6){\line(2,0){.6}}
\put(10.2,.3){\line(0,-3){.9}}
\put(10.5,.3){\line(0,-3){.9}}
\put(10.8,.3){\line(0,-3){.9}}
\put(10.4, -1){$\rho_6$}
\put(2.5,-1.6){$\oplus$}
\put(2.9,-1.6){\line(4,0){1.2}}
\put(2.9,-1.3){\line(4,0){1.2}}
\multiput(2.9,-1.6)(.3,0){5}{\line(0,1){.3}}
\put(2.9,-1.6){\line(0,-2){.6}}
\put(2.9,-1.9){\line(1,0){.3}}
\put(2.9,-2.2){\line(1,0){.3}}
\put(3.2,-1.6){\line(0,-2){.6}}
\put(3.3, -2.8){$\rho_7$}
\put(4.2,-1.6){$\oplus$}
\put(4.8,-1.3){\line(2,0){.6}}
\put(4.8,-1.6){\line(2,0){.6}}
\put(4.8,-1.9){\line(2,0){.6}}
\put(4.8,-1.3){\line(0,-4){1.2}}
\put(5.1,-1.3){\line(0,-4){1.2}}
\put(5.4,-1.3){\line(0,-2){.6}}
\put(4.8,-2.2){\line(1,0){.3}}
\put(4.8,-2.5){\line(1,0){.3}}
\put(5,-2.8){$\rho_8$}
\end{picture}$$
\vskip2cm
\noindent
We will write the quantum dimensions
for the above representations and the braiding
eigenvalues for parallel strands which will be 
useful for computation of new polynomial
invariants in topological string theory.
\begin{eqnarray}
{\rm dim}_q \rho_1=c {[N+2][N-2][N-3] \over ~[6][2]^2}
&;& \lambda_{\rho_1} ^{(+)}(R, R) = a q^{-3 (N^3+N^2-8N-12)
\over2 N (N+2)}\label{rone}\\
{\rm dim}_q \rho_2= c {[N][N+2][N-2]\over [3] [5]}&;&
\lambda_{\rho_2}^{(+)}(R,R)=a q^{- 3 (N^2-6) \over 2N}\label{rtwo}\\
{\rm dim}_q \rho_3= c {[N][N+2][N-2]\over [3] [5]}&;&
\lambda_{\rho_3}^{(+)}(R,R)=-a q^{- 3 (N^2-6) \over 2N}\label{rthree}\\
{\rm dim}_q \rho_4 = c {[N][N+2][N+3] [3]\over [2]^2 [5] [4]}&;&
\lambda_{\rho_4}^{(+)} (R,R)= a q^{-(3N^2+5N-18) \over 2N}\label{rfour}\\
{\rm dim}_q \rho_5 =c { [N][N+1][N+2] \over [2]^2 [4] [3]}&;&
\lambda_{\rho_5}^{(+)} (R,R) = -a q^{-3 (N+3) (N-2) \over 2N}\label{rfive}\\ 
{\rm dim}_q \rho_6 = c {[N][N-1][N-2] \over [2]^2 [3] [4]}&;&
\lambda_{\rho_6}^{(+)} (R,R)= a q^{-3 (N-3) (N+2) \over 2N}\label{rsix}\\
{\rm dim}_q \rho_7 = c {[N+2][N+3][N-2] \over [6] [2]^2}&;&
\lambda_{\rho_7}^{(+)}= -a q^{-3 (N^4+6N^3+5N^2-24N-36) \over 2N (N+2)
(N+3)}\label{rseven}\\
{\rm dim}_q \rho_8= c {[3][N][N-2][N-3] \over [2]^2 [4] [5]}&;&
\lambda_{\rho_8}^{(+)}= -a q^{-(3N^2 - 5N -18) \over 2N}
\label{reight}
\end{eqnarray}
where $c = [N][N+1][N-1]/ [3]$ and $a=q^{6 (N^2-3)/ 2N}$. The $\pm$
signs in the braiding eigenvalues are deduced from topological
equivalence $L_{\pm 1} \equiv {\rm unknot}$.
 
For the anti-parallelly oriented strands carrying symmetric
representation $R_n$, the braiding eigenvalue (\ref {aeval})
is given by
\be
\lambda^{(-)}_{{\tilde \rho}_{\ell}} (R_n, \bar{R}_n) \ = \ (-)^\ell~q^{(N-1)\ell/2} q^{\ell^2/2}
~~;~~\ell = 0,1,\ldots , n~.
\ee
Here these eigenvalues for $\ell = 0,1,2,\ldots n$ correspond
respectively to the irreducible representations in the product $R_n \otimes
\bar{R}_n = \oplus \tilde {\rho}_\ell$ :
$$
\begin{picture}(10,1.5)
\put(0,0){\line(1,0){1.8}}
\put(0,.3){\line(1,0){1.8}}
\multiput(0,0)(.3,0){7}{\line(0,1){.3}}
\multiput(.15,.15)(.3,0){6}{$.$}
\put(.7,.5){\vector(-1,0){.7}}
\put(1.1,.5){\vector(1,0){.7}}
\put(.8,.5){$n$}
\put(2,0){$\otimes$}
\put(.8,-.4){$\bar R_n$}
\put(2.6,0){\line(1,0){1.8}}
\put(2.6,.3){\line(1,0){1.8}}
\multiput(2.6,0)(.3,0){7}{\line(0,1){.3}}
\put(3.3,.5){\vector(-1,0){.7}}
\put(3.7,.5){\vector(1,0){.7}}
\put(3.4,.5){$n$}
\put(3.4,-.4){$R_n$}
\put(4.8,0){$=$}
\put(5.4,0){$\oplus_{\ell=0}^n$}
\put(6.5,0){\line(1,0){2.4}}
\put(6.5,.3){\line(1,0){2.4}}
\multiput(6.5,0)(.3,0){9}{\line(0,1){.3}}
\multiput(6.65,.15)(.3,0){4}{$.$}
\put(7.5,-.4){${\tilde \rho}_{\ell}$}
\put(6.8,.5){\vector(-1,0){.3}}
\put(7.4,.5){\vector(1,0){.3}}
\put(7,.5){$\ell$}
\put(8,.5){\vector(-1,0){.3}}
\put(8.6,.5){\vector(1,0){.3}}
\put(8.2,.5){$\ell$}
\end{picture}
$$
\noindent Here  boxes with dot represents a column of length $N-1$.
The quantum dimension for $\tilde {\rho}_{\ell}$ (\ref {qdim})
is:
\begin{equation}
{\rm dim}_q  {\tilde {\rho}}_{\ell} = {([N][N+1] \ldots [N+2\ell-1])
\prod_{a=1}^{N-2} ([N-a] [N-a+1] \ldots [N-a+\ell-1])
\over [\ell]! ([N+2 \ell-2] [N+2 \ell-3] \ldots [N+\ell-1])  
\prod_{a=1}^{N-2} ([a][a+1] \ldots [a + \ell-1])}
\end{equation}

Similarly, for antisymmetric representations $\hat R_n$, we
have the following eigenvalues in anti-parallely oriented strands:
\begin{equation}
\lambda_{\hat {\tilde \rho}_{\ell}}^{(-)}(\hat R_n, \bar {\hat R_n})~=~
(-1)^{\ell} q^{(N-1) \ell \over 2} q^{\ell (2 - \ell) \over 2}
\end{equation}
where the representations ${\tilde {\hat \rho}_{\ell}}$ appear
in the following tensor product:
\vskip.5cm
$$\begin{picture}(10,1.5)
\put(0,0){\line(0,1){1.8}}
\put(.3,0){\line(0,1){1.8}}
\multiput(0,0)(0,.3){7}{\line(1,0){.3}}
\put(.5,.7){\vector(0,-1){.7}}
\put(.5,1.1){\vector(0,1){.7}}
\put(.5,.8){$n$}
\put(1,.8){$\otimes$}
\put(-.6,.8){${\hat R}_n$}
\put(2.3,0){\line(0,1){2.1}}
\put(2.6,0){\line(0,1){2.1}}
\multiput(2.3,0)(0,0.3){8}{\line(1,0){.3}}
\put(3.1,.7){\vector(0,-1){.7}}
\put(3.1,1.1){\vector(0,1){.7}}
\put(3.1,.8){$N-n$}
\put(1.7,.8){${\bar {\hat R}}_n$}
\put(4.4,.6){$=$}
\put(5,.6){$\oplus_{\ell=0}^n$}
\put(6.8,2){\line(0,-1){3}}
\put(7.1,2){\line(0,-1){1.8}}
\put(6.5,-1){\line(0,1){3}}
\put(6.5,2){\line(2,0){.6}}
\put(6.5,1.7){\line(2,0){.6}}
\put(6.5,1.4){\line(2,0){.6}}
\put(6.5,1.1){\line(2,0){.6}}
\put(6.5,.8){\line(2,0){.6}}
\put(6.5,.5){\line(2,0){.6}}
\put(6.5,.2){\line(2,0){.6}}
\put(6.5,-.1){\line(1,0){.3}}
\put(6.5,-.4){\line(1,0){.3}}
\put(6.5,-.7){\line(1,0){.3}}
\put(6.5,-.7){\line(1,0){.3}}
\put(6.5,-1){\line(1,0){.3}}
\put(6.7,-1.5){${\tilde {\hat \rho}}_{\ell}$}
\put(7.2,1.2){\vector(0,1){.9}}
\put(7.2,1){$\ell$}
\put(7.2,.8){\vector(0,-1){.65}}
\put(7.2,-.6){\vector(0,-1){.5}}
\put(7.2,-.2){\vector(0,1){.5}}
\put(7.2,-.4){$N-2\ell$}
\end{picture}~.$$
\vskip.8cm
and the quantum dimension for ${\tilde {\hat \rho}_{\ell}}$ is
\begin{equation}
{\rm dim}_q {\tilde {\hat \rho}_{\ell}}= {([N][N-1] \ldots [\ell+1])
([N+1] [N] \ldots [N-\ell+2]) \over [N-2 \ell]! [\ell]! ([N-2\ell+2]
[N-2\ell+3] \ldots [N-\ell+1])}
\end{equation}

We will list the knot invariant formula for some knots (carrying
arbitrary $SU(N)$ representation $R$) as shown in Fig.~4.

\myfigure{\epsfxsize=4in \epsfbox{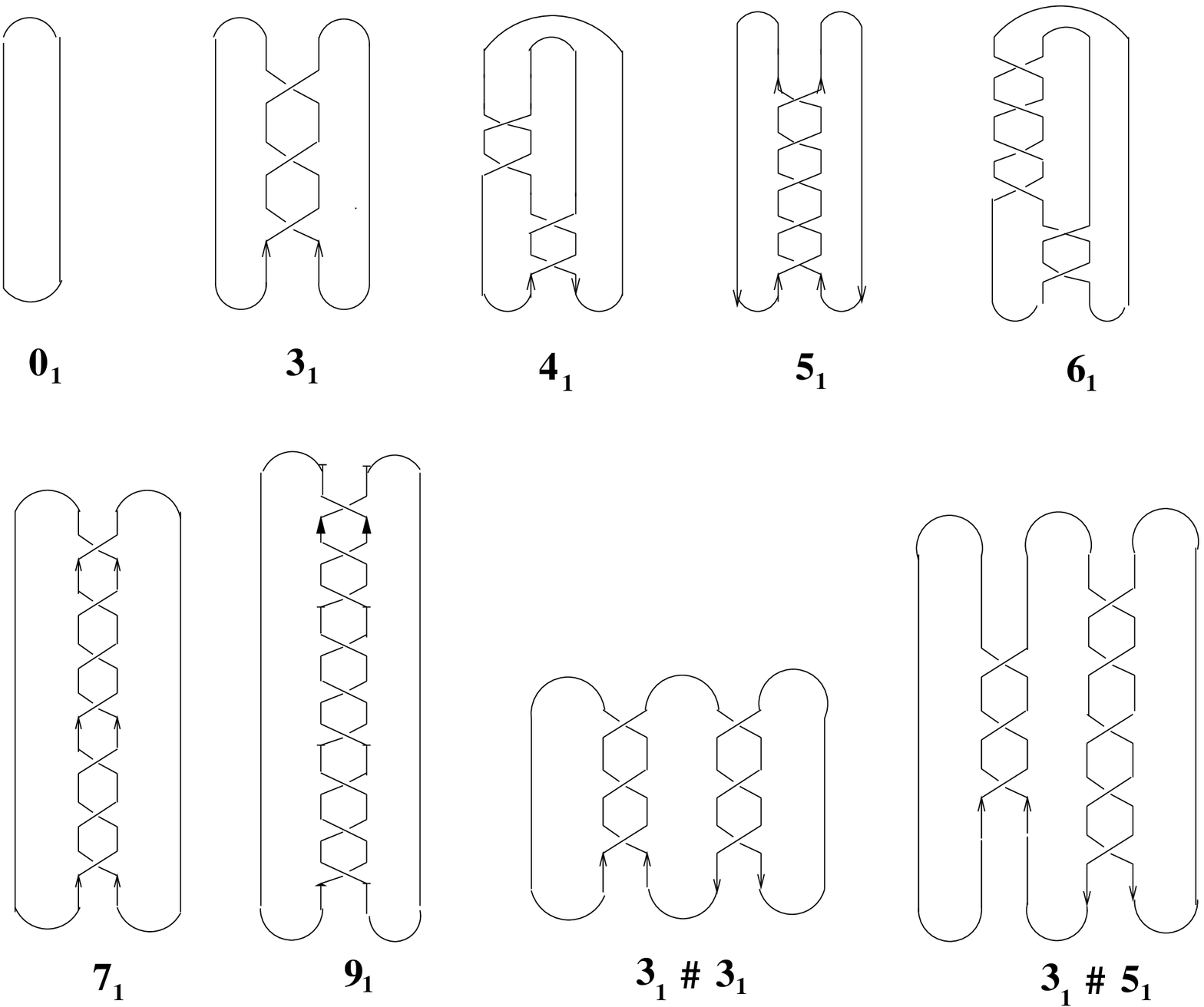}} {\sl Some knots upto
9 crossings obtained from braids.} 
\begin{eqnarray}
V_R[0_1]&=&dim_qR\label{3.31}\\
V_R[3_1]&=&\sum_{\rho_{\ell} \in R \otimes R} dim_q\rho_{\ell} 
\left(\lambda_{\rho_{\ell}}^{(+)} (R,R)\right)^3\label{3.32}\\
V_R[4_1]&=&
\sum_{\rho_{\ell},\rho_j \in R \otimes \bar R}
\sqrt{dim_q\rho_j dim_q\rho_\ell}~
a_{\rho_j\rho_\ell}\left(\matrix {R &\bar R \cr R &\bar R\cr}\right)
(\lambda_{\rho_j}^{(-)} (R,R))^2 \nonumber\\ 
~&~&~~~~~~~~~~(\lambda_{\rho_{\ell}}^{(-)} (R,R))^{-2}\label{3.33}\\ 
V_R[5_1]&=&\sum_{\rho_{\ell} \in R \otimes R} dim_q\rho_{\ell} 
\left(\lambda_{\rho_{\ell}}^{(+)} (R,R)\right)^5\label{3.34}\\
V_R[6_1]&=&\sum_{\rho_\ell,\rho_j \in R \otimes \bar R}
\sqrt{dim_q\rho_jdim_q\rho_\ell}~
a_{\rho_j\rho_\ell}\left(\matrix{ R & \bar R \cr R & \bar R \cr} \right) 
(\lambda_{\rho_j}^{(-)} (R, \bar R) )^4 \nonumber\\ 
~&~&~~~~~~~~(\lambda_{\rho_\ell}^{(-)} (R, \bar R) )^{-2}\label{3.35}\\ 
V_R[7_1]&=&\sum_{\rho_{\ell} \in R \otimes R} dim_q\rho_{\ell} 
\left(\lambda_{\rho_{\ell}}^{(+)} (R,R)\right)^7\label{3.36}\\
V_R[9_1]&=&\sum_{\rho_{\ell} \in R \otimes R} dim_q\rho_{\ell} 
\left(\lambda_{\rho_{\ell}}^{(+)} (R,R)\right)^9\label{3.37}\\
V_R[3_1 \# 3_1]&=& {(V_R [3_1])^2 \over V_R[0_1]}\label{3.38}\\
V_R[3_1 \# 5_1]&=& {V_R [3_1] V_R[5_1] \over V_R[0_1]}\label{3.39}
\end{eqnarray}
Similarly, the invariants for some two-component links (see Fig.~5) are:
\begin{eqnarray}
V_{R_1,R_2}[0_1^2]&=& {\rm dim}_q R_1 {\rm dim}_q R_2 \label{3.40}\\
V_{R_1,R_2}[2_1^2]&=& \sum_{\rho_\ell \in R_1 \otimes R_2} {\rm dim}_q \rho_\ell
(\lambda_{\rho_\ell}^{(+)} (R_1,R_2))^2\label{3.41}\\
V_{R_1,R_2}[4_1^2]&=&\sum_{\rho_\ell \in R_1 \otimes R_2} {\rm dim}_q \rho_\ell
(\lambda_{\rho_\ell}^{(+)} (R_1,R_2))^4\label{3.42}\\
V_{R_1,R_2}[6_1^2]&=&\sum_{\rho_\ell \in R_1 \otimes R_2} {\rm dim}_q \rho_\ell
(\lambda_{\rho_\ell}^{(+)} (R_1,R_2))^6\label{3.43}
\end{eqnarray}
\myfigure{\epsfxsize=2.5in \epsfbox{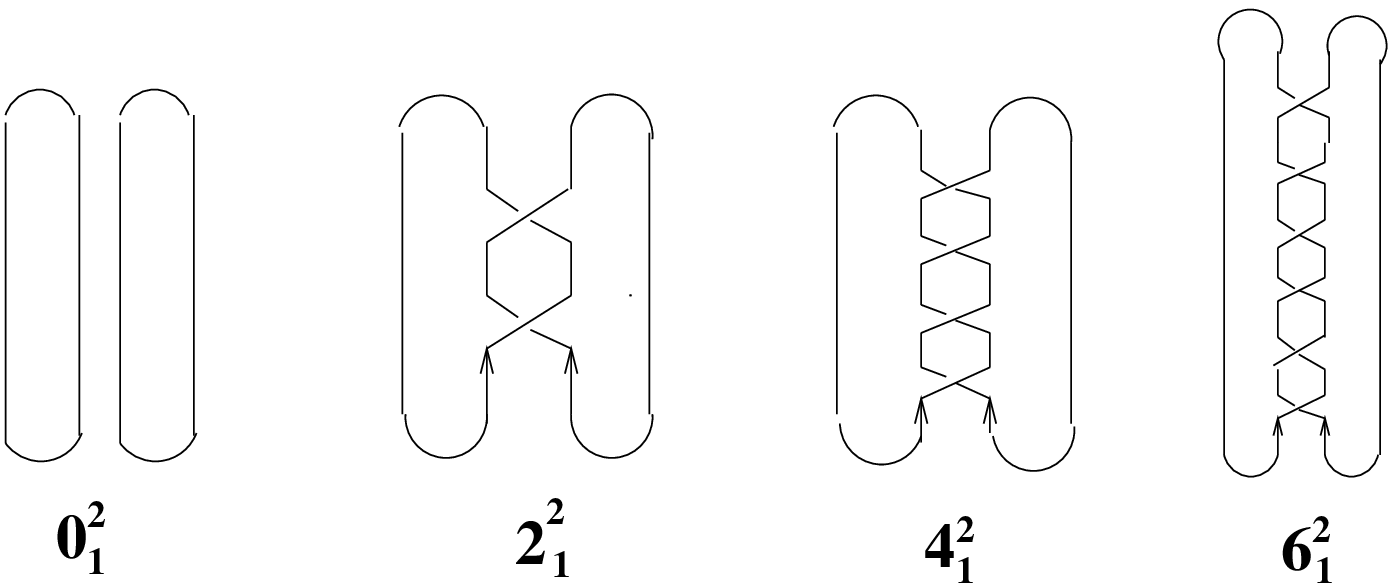}} {\sl Some two-component links 
obtained from braids.} 
It is appropriate to stress that there is no formula for $SU(N)_q$ 
quantum Racah coefficients in the literature. We have used 
topological equivalence of knots in deriving  some of them but not
all \cite{gkr}. 
For example, the form of the duality matrix for $R= R_2 {\rm ~or~} 
\hat R_2$ (horizontal two box or verical 
two box representation), we have
\be
a_{\rho_i \rho_j} \left( \matrix{ R &\bar R \cr
R & \bar R\cr} \right)
 =  \frac{1}{dim_q R} \pmatrix { \sqrt{dim_q
\rho_0} & \sqrt{ dim_q \rho_1} & \sqrt{dim_q \rho_2} \cr
\sqrt{dim_q \rho_1} & \frac{ dim_q \rho_2}{dim_q R-1} -1 &
\frac{-\sqrt{dim_q \rho_1 dim_q \rho_2}}{dim_q R - 1} \cr
\sqrt{dim_q \rho}_2 & - \frac{\sqrt{dim_q \rho}_1 dim_q
\rho_2}{dim_q R -1} & {\frac{dim_q \rho_1 }{dim_q
R - 1}-1} }~,
\ee
where $\rho_i \in R \otimes \bar R$.

\section{Explicit Computation of $f_R$} 
\subsection{Torus Knots of the Type $(2,2m+1)$ Upto Nine Crossings}
In this section, we present explicit results on the computation 
of $f_R$ for knots that consist of two stranded braids with 
upto nine crossings for representations of $SU(N)$ upto three boxes
in the Young tableau. We will show that the Ooguri-Vafa prediction 
of eq. (\ref{lmaeq}) is exactly reproduced, providing strong
evidence for the conjecture in \cite{ov}. Essentially, $f_R$ is 
calculated from a knowledge of $\langle{\mbox{Tr}}_RU\rangle$ with
the methods described in the last section, and by using 
eqs. (\ref{onebox}), (\ref{twoboxes}), (\ref{threeboxes}). The
evaluation of the right handed trefoil knot was already done
in \cite{lm}, but we will state the results here for the sake
of completeness. Let us therefore begin with the case of the 
right handed trefoil knot, also known as the $3_1$ knot. The single
box representation gives us from (\ref{3.32}) 
\begin{equation}
f_{\one}\left(3_1\right)=\langle{\mbox{Tr}}_{\one}U\rangle=-\denom
\left[q^{-1}\lambda^{\frac{1}{2}}\left(-1+\lambda\right)
\left(-1-q^2+q\lambda\right)\right] 
\end{equation}
Using this, we can calculate the expressions for the 
representations $\twohor$ and $\twover$ (\ref{twoboxes}). 
The results are
\begin{eqnarray}
f_{\twohor}\left(3_1\right)&=&\denom
\left[\lambda q^{-{1\over 2}}\left(1+ q^2\right){\left(-1+\lambda\right) }^2
\left(q-\lambda- q^2\lambda+q\lambda^2\right)\right] 
\nonumber\\
f_{\twover}\left(3_1\right)&=&\denom
\left[\lambda q^{-{3\over 2}}\left(1+q^2\right){\left(-1+\lambda\right)}^2
\left(-q+\lambda+q^2\lambda-q\lambda^2\right)\right]
\end{eqnarray}
We now list the results for representations involving three boxes in their
Young tableau (\ref{threeboxes}). The explicit results for these, 
(which have also been tabulated in \cite{lm}) 
using eq. (\ref{3.32}) are
\begin{eqnarray}
f_{\threehor}\left(3_1\right)&=&\denom
\left[q^{-1}\lambda^3\left(q-\lambda\right)
{\left(-1+\lambda\right)}^2\left(-1+q\lambda\right)
\left\{q^2\left(1+q+q^3\right)+\right.\right.\nonumber\\&~&
\left.\left.q^2\lambda^2\left(1+q^2\right)
\left(1+q+q^4\right)-\lambda\left(1+q+q^3\right)
\left(1+q+q^2+q^3+q^4\right)\right\}\right] 
\nonumber\\
f_{\threever}\left(3_1\right)&=&\denomneg
\left[q^{-9}\lambda^{3\over 2}{\left(-1+\lambda\right)}^2
\left( -q + \lambda \right)\left( -1 + q\lambda \right)
\left\{ q^3 + q^5 + q^6 +\right.\right.\nonumber\\&~&\left.\left.
\lambda^2\left( 1 + q^2 \right)\left(1+q^3+q^4\right) -
q\lambda\left(1+q^2+q^3\right)
\left(1+q+q^2+q^3+q^4\right)\right\}\right] 
\nonumber\\
f_{\mixed}\left(3_1\right)&=&
\denom
\left[q^{-\frac{19}{2}}\lambda^{3\over 2}\left(1+q+q^2 \right)
\left\{q^{\frac{17}{2}}+ q^{\frac{15}{2}}\lambda^6
\left(1+q^2\right)\right. \right.\nonumber\\
&~&\left. \left.- q^{\frac{13}{2}}\lambda{\left(1+q+q^2\right)}^2
- q^{\frac{11}{2}}\lambda^3\left(1+q^2\right)
\left(1+q+q^2 \right) 
\left(2+q\left(3+2q \right)\right)\right.\right.\nonumber\\
&~&\left.\left.
+q^{\frac{11}{2}}\lambda^4\left( 1 + q + q^2 \right)
\left(1+ q\left( 4 + q \right)\left( 1 + q^2 \right)
\right) \right.\right.\nonumber\\
&~&\left.\left.
+q^{\frac{11}{2}}\lambda^2\left(
1+q^2\right)
\left(1+3q+6q^2+3q^3+q^4\right)
\right.\right.\nonumber\\&~&\left.\left.
-q^{\frac{13}{2}}\lambda^5\left(
2+3q+3q^2+3q^3+2q^4\right)\right\}\right]
\end{eqnarray}
Let us now list the invariants for the knot denoted by $5_1$. In this 
case from eq. (\ref{3.34}) we will evaluate, as before, the invariants 
in representations with upto three boxes in the Young diagram. For the 
representation $R=\one$, we obtain 
\begin{equation}
f_{\one}\left(5_1\right)=\denomneg
\left[q^{-2}\lambda^{3\over 2}
\left(-1+ \lambda^{\frac{1}{2}}\right)
\left(1+\lambda^{\frac{1}{2}}\right)
\left(-1-q^2-q^4+\lambda q+\lambda q^3\right) 
\right] 
\end{equation}
And for $R=\twohor {\mbox{and}}~\twover$, we get, from (\ref{3.34}) and
(\ref{twoboxes}),
\begin{eqnarray}
f_{\twohor}\left(5_1\right)&=&\denom
\left[q^{-\frac{5}{2}}\lambda^3\left(1+q^2\right)
\left(1+q^2+q^3+q^4+q^6\right)
{\left(\lambda-1\right)}^2\right.\nonumber\\&~&~~~~~~~~~~~~~\left.
\times\left(q-\lambda-\lambda q^2+\lambda^2 q\right)\right]
\nonumber\\
f_{\twover}\left(5_1\right)&=&\denomneg
\left[q^{-\frac{15}{2}}\lambda^3\left(1+q^2\right)
\left(1+q^2+q^3+q^4+q^6\right)
{\left(\lambda-1\right)}^2\right.\nonumber\\&~&~~~~~~~~~~~~~\left.
\times\left(q-\lambda-\lambda q^2+\lambda^2 q\right)\right]
\end{eqnarray}
For representations with three boxes in the Young tableau, we get 
from eqs. (\ref{3.34}) and (\ref{threeboxes}),
\begin{eqnarray}
f_{\threehor}\left(5_1\right)&=&\denomneg
\left[q^{-4}\lambda^{9\over 2}
\left(1+q\left(q-1\right)\right)
{\left(\lambda-1\right)}^2\left(\lambda-q\right)
\left(\lambda q-1\right)\right.\nonumber\\&~&\left.
\times\left\{q^2\left(1+q^2\right)\left(1+q+q^4\right)
\left(1+2q+2q^2+2q^3+q^4+q^5+q^6+q^7\right)
\right.\right.\nonumber\\
&~&\left.\left.+q^2\lambda^2
\left(1+q+q^2+q^3+q^4+q^5+q^6\right)\right.\right.\nonumber\\
&~&\left.\left.
\left(1+2q+2q^2+2q^3+2q^4+q^6+q^7+q^8+q^{10}\right)
\right.\right.\nonumber\\
&~&\left.\left.-\lambda\left(
1+3q+5q^2+8q^3+12q^4+16q^5+20q^6+23q^7 + 
22q^8+19q^9
\right.\right.\right.\nonumber\\&~&\left.\left.\left.
+16q^{10}+13q^{11}+10q^{12}+8q^{13} + 
5q^{14}+3q^{15}+2q^{16}+q^{17}\right)\right\}\right]
\nonumber\\
f_{\threever}\left(5_1\right)&=&\denomneg
\left[q^{-18}\lambda^{9\over 2}
\left( 1 + q\left(q-1\right)\right) 
{\left(\lambda-1\right)}^2\left(\lambda-q\right) 
\left( \lambda q -1\right)\right.\nonumber\\&~&\left.
\left\{ q^3\left( 1 + q^2 \right) \left( 1 + q^3 + q^4 \right) 
\left( 1 + q\left( 1 + q \right) 
\left( 1 + q^2 + q^3 + q^4 + q^5 \right) \right)
\right.\right.\nonumber\\ &~&\left.\left.
+\lambda^2\left(1+q+2q^2+3q^3+4q^4+4q^5+6q^6+7q^7+9q^8+ 
10q^9+10q^{10}\right.\right.\right.\nonumber\\&~&\left.\left.\left.
+9q^{11}+9q^{12}+7q^{13}+5q^{14}+3q^{15}+q^{16}\right)
\right.\right.\nonumber\\&~&\left.\left.-\lambda q
\left(1+2q+3q^2+5q^3+8q^4+10q^5+13q^6+16q^7 + 
19q^8+22q^9\right.\right.\right.\nonumber\\&~&\left.\left.\left.
+23q^{10}+20q^{11}+16q^{12}+12q^{13} + 
8q^{14}+5q^{15}+3q^{16}+q^{17}\right)\right\}\right]
\nonumber\\
f_{\mixed}\left(5_1\right)&=&\denom
\left[q^{-9}\lambda^{9\over 2}
\left(1+q\left(q-1\right)\right)
{\left(\lambda-1\right)}^2\left(\lambda-q\right)
\left(\lambda q-1\right)\right.\nonumber\\
&~&\left.
\left\{\lambda^2 q\left(1+q+q^2+q^3+q^4+q^5+q^6
\right)\left(1+q\left(1+q+q^2\right) 
\left(1+2q+q^3\right)\right)\right.\right.\nonumber\\
&~&\left.\left. 
+q^2\left(1+q^2\right)\left(1+q+q^2\right) 
\left(1+2q+2q^2+q^3+2q^4+2q^5+q^6\right)\right.\right.\nonumber\\
&~&\left.\left.-\lambda\left(
1+3q+6q^2+10q^3+15q^4+20q^5+25q^6+27q^7+ 
25q^8+20q^9\right.\right.\right.\nonumber\\&~&\left.\left.\left.
+15q^{10}+10q^{11}+6q^{12}+3q^{13}+q^{14}\right)\right\}\right]
\end{eqnarray}
Let us now consider the knot $7_1$, i.e the two stranded braid with 
seven crossings, the invariant in the defining representation of
$SU(N)$ is, from eq. (\ref{3.36}), 
\begin{eqnarray}
f_{\one}\left(7_1\right)&=&\denomneg
\left[q^{-3}\lambda^{1\over 2}
\left(\lambda^{1\over 2}-1\right) 
\left(\lambda^{1\over 2}+1\right)\right.\nonumber\\
&~&\left. 
\left(-1-q^2-q^4-q^6+\lambda q+\lambda q^3+\lambda q^5 
\right)\right] 
\end{eqnarray}
Similarly, we obtain from (\ref{3.36}) and (\ref{twoboxes}),
\begin{eqnarray}
f_{\twohor}\left(7_1\right)&=&
\denom
\left[q^{-\frac{9}{2}}\lambda^5\left(1+q^2\right)
\left(1+q^2+q^3+2q^4+q^5+2q^6+q^7+2q^8\right. \right. \nonumber\\
&~&\left. \left.+q^9+
q^{10} +q^{12}\right){\left(\lambda-1\right)}^2
\left(q-\lambda-\lambda q^2+\lambda^2 q\right)\right]
\nonumber\\
f_{\twover}\left(7_1\right)&=&
\denomneg
\left[q^{-\frac{23}{2}}\lambda^5\left(1+q^2\right)
\left(1+q^2+q^3+2q^4+q^5+2q^6\right.\right.
\nonumber\\&~&\left.\left.+q^7+2q^8+q^9 +
q^{10}+q^{12}\right){\left(\lambda-1\right)}^2
\left(q-\lambda-\lambda q^2+\lambda^2 q\right)\right]
\end{eqnarray}
The expressions for the invariants in representations with three 
boxes in the Young diagram becomes highly 
complicated as we increase the number of crossings, and we relegate
the details of these expressions for the present case of 
the knot $7_1$ to the appendix. Before we close this section, let us also
present the results for the two stranded braid with nine crossings.
We will do so for the representations $\one$, $\twohor$ and $\twover$. 
The expressions for the invariants with higher number of boxes in the
Young diagram will not be presented here for reasons of space. 
However, we hasten to assure the reader that in all these cases, we
have found perfect agreement with the Ooguri-Vafa conjecture of
eq. (\ref{lmaeq}).
The explicit results for invariants of the knot $9_1$ are, from
eqs. (\ref{3.37}) and (\ref{twoboxes}), 
\begin{eqnarray}
f_{\one}\left(9_1\right)&=&
\denom\left[
q^{-4}\lambda^{7\over 2}
\left(\lambda-1\right)\right.\nonumber\\
&~&\left.\left\{-1-q^2-q^4-q^6-q^8+\lambda\left(q+q^3+ 
q^5+q^7\right)\right\}\right]
\end{eqnarray}
in the defining representation, and for two boxes in the Young diagram, 
we obtain
\begin{eqnarray}
f_{\twohor}\left(9_1\right)&=&
\denom
\left[q^{-\frac{13}{2}}\lambda^7\left(1+q^2\right)
\left(1+q^2+q^3
\right.\right.\nonumber\\&~&\left.\left.
+2q^4+q^5+3q^6+2q^7+3q^8+2q^9+3q^{10}+2q^{11}+3q^{12}+q^{13}
\right.\right.\nonumber\\
&~&\left.\left.+2q^{14}+q^{15}+q^{16}+q^{18}\right)
{\left(\lambda-1\right)}^2
\left(q-\lambda-\lambda q^2+\lambda^2 q\right)\right]
\nonumber\\
f_{\twover}\left(9_1\right)&=&
\denomneg
\left[q^{-\frac{31}{2}}\lambda^7\left(1+q^2\right)
\left(1+q^2+q^3\right.\right.\nonumber\\
&~&\left.\left.+2q^4+q^5+3q^6+2q^7+3q^8 +
2q^9+3q^{10}+2q^{11}+3q^{12}+q^{13}
\right.\right.\nonumber\\
&~&\left.\left.+2q^{14} +
q^{15}+q^{16}+q^{18}\right){\left(\lambda-1\right)}^2
\left(q-\lambda-\lambda q^2+\lambda^2 q\right)\right]
\end{eqnarray}

\subsection{The Knots $4_1$ and $6_1$}
In this subsection, we go beyond torus knots and compute the 
invariants for the knots
$4_1$ and $6_1$ using the methods described in the last section. 
We will do so for representations containing one and two boxes in the
Young tableau. For the knot $4_1$, we obtain from eq. (\ref{3.33}) 
following
\begin{equation}
f_{\one}\left(4_1\right)=\denom  
\left[q^{-1}\lambda^{-\frac{3}{2}}\left(\lambda-1\right) 
\left(q+\lambda^2q-\lambda
\left(1-q+q^2\right)\right)\right] 
\end{equation}
For representations consisting of two boxes in the Young diagram, 
we get the following expressions from eqs. (\ref{3.33}) and 
(\ref{twoboxes})
\begin{eqnarray}
f_{\twohor}\left(4_1\right)&=&\denom
\left[q^{-\frac{5}{2}}\lambda^{-3}{\left(\lambda-1\right)}^2
\left(\lambda-q\right)
\left(\lambda^2 q^2-1\right)
\left(\lambda q^2-1\right)\right]\nonumber\\
f_{\twover}\left(4_1\right)&=&\denomneg
\left[q^{-\frac{5}{2}}\lambda^{-3}{\left(\lambda-1\right)}^2
\left(\lambda-q^2\right)\right.\nonumber\\
&~&\left.
\times\left(\lambda^2-q^2\right)\left(\lambda q-1\right)\right] 
\end{eqnarray}
Similar expressions can be calculated for the knot $6_1$. We obtain, in 
this case, from eqs. (\ref{3.35}) and (\ref{twoboxes}),
the following result for $R=\one$ 
\begin{eqnarray}
f_{\one}\left(6_1\right)&=&\denom\left[q^{-1}\lambda^{-{3\over 2}}
\left\{-q+\lambda^2 q+\lambda^4 q + 
\lambda\left(1-q+q^2\right)- 
\right.\right.\nonumber\\&~&\left.\left.
\lambda^3\left(1+q^2\right)\right\}\right] 
\end{eqnarray}
and the representations $\twohor$ and $\twover$, we obtain, in a similar
fashion as before,
\begin{eqnarray}
f_{\twohor}\left(6_1\right)&=&\denom
\left[q^{-\frac{5}{2}}\lambda^{-3}{\left(\lambda-1\right)}^2
\left(\lambda-q\right) 
\left(\lambda q-1\right)\right.\nonumber\\&~&\left.
\left(-1-\lambda-2\lambda q+\lambda q^2-2\lambda^2q+\lambda^2q^3- 
\lambda^3 q+\lambda^3q^4+\lambda^4q^3+\lambda^4q^5\right)\right]
\nonumber\\
f_{\twover}\left(6_1\right)&=&\denomneg
\left[q^{-\frac{9}{2}}\lambda^{-3}{\left(\lambda-1\right)}^2
\left(\lambda-q\right)\left(\lambda q-1\right)
\right.\nonumber\\&~&\left.
\left\{-q^5+\lambda^2q^2\left(1-2q^2\right)
+\lambda^4\left(1+q^2\right)- 
\lambda^3q\left(q^3-1\right) 
\right.\right.\nonumber\\&~&\left.\left.
+\lambda q^3\left(1-2q-q^2\right)\right\}\right] 
\end{eqnarray}
These results are again in full agreement with the conjecture of Ooguri and 
Vafa \cite{ov} thus providing further strong evidence in support of 
their arguments.
\subsection{Connected Sums}
We will now calculate the new invariants obtained from knots that
are connected sums. The knots $K=m_1\#n_1$ where $m$ and $n$ are
odd integers are of this form. We will explicitly evaluate the new invariants
for the knots $K_1=3_1\#3_1 {\mbox{and}} K_2=3_1\#5_1$ (as drawn in 
figure 4.) for 
representations upto two boxes in the Young diagram, although from our 
methods, these can be easliy extended to general $m$ and $n$.
Let us start with the knot $K_1=3_1\#3_1$. For the defining 
representation, the invariant is given by eq. (\ref{3.38})
\begin{equation}
f_{\one}(3_1\#3_1)=\denom
\left[q^{-2}\lambda^{3\over 2}\left(\lambda-1\right)
{\left(1+q^2-q\lambda\right)}^2\right]
\end{equation}
For the representation $R={\twohor}$, we obtain from 
eqs. (\ref{3.38}) and (\ref{twoboxes}), 
\begin{eqnarray}
f_{\twohor}(3_1\#3_1)&=&\denom
\left[q^{-7}\lambda^3
\left\{q^{\frac{17}{2}}\lambda^6\left(1+q^2+q^4\right)- 
q^{\frac{13}{2}}\lambda^5
\left(2+3q+5q^2\right.\right.\right.\nonumber\\
&~&\left.\left.\left.+5q^3+5q^4+3q^5+3q^6+2q^7\right)
+q^{\frac{11}{2}}
\left(2+4q^2+2q^3+3q^4\right. \right. \right. \nonumber\\
&~&\left. \left. \left. +2q^5+3q^6+q^8 \right)
-q^{\frac{9}{2}}\lambda\left(2+4q
+8q^2+12q^3+14q^4 \right. \right. \right. \nonumber\\
&~&\left. \left. \left. +13q^5+13q^6+9q^7 + 
5q^8+3q^9+q^{10}\right)
-q^{\frac{9}{2}}\lambda^3 \left(2+8q+14q^2 \right. \right. \right. \nonumber\\
&~&\left. \left. \left. +25q^3+29q^4+30q^5+26q^6+20q^7+ 
12q^8+7q^9+3q^{10}\right)
\right.\right.\nonumber\\
&~&\left.\left. 
+q^{\frac{11}{2}}\lambda^4
\left(4+6q+15q^2
+15q^3+19q^4+14q^5+13q^6+6q^7  
+\right.\right.\right.\nonumber\\&~&\left.\left.\left.
+ 6q^8+q^9\right)+q^{\frac{9}{2}}\lambda^2
\left(4+6q+18q^2+21q^3+30q^4+26q^5 \right. \right. \right. \nonumber\\
&~&\left. \left. \left. +27q^6 +16q^7+ 
13q^8+5q^9+3q^{10}\right)\right\}\right]
\end{eqnarray}
and similarly for $R={\twover}$, we obtain 
\begin{eqnarray}
f_{\twover}(3_1\#3_1)&=&\denomneg
\left[q^{-{15\over 2}}\lambda^3{\left(\lambda-1\right)}^2
\left(\lambda-q\right)\left(q\lambda-1\right)
\right.\nonumber\\&~&\left.
\left\{1+3q^2+2q^3+3q^4+2q^5+4q^6+2q^8-\lambda\left(1+ 
q+q^2+3q^3+\right.\right.\right.\nonumber\\&~&\left.\left.\left.
3q^4+3q^5+2q^6+2q^7\right)+\lambda^2\left(q+q^3 + 
q^5\right)\right\}\right]
\end{eqnarray}
Next, let us consider the knot $3_1\#5_1$. Using eqs. (\ref{3.39}) 
and (\ref{twoboxes}), we get the following results: for $R={\one}$,
the invariant polynomial is given by
\begin{equation}
f_{\one}(3_1\#5_1)
\denom
\left[q^{-3}\lambda^{5\over 2}\left(\lambda-1\right)
\left(-1-q^2+q\lambda\right) 
\left(-1-q^2-q^4+\lambda q+\lambda q^3\right)\right]
\end{equation}
for $R={\twohor}$, we get
\begin{eqnarray}
f_{\twohor}(3_1\#5_1)&=&\denom
\left[q^{-9}\lambda^5
\left\{\lambda^6q^{\frac{17}{2}}
\left(1+q+2q^2+q^3+3q^4+q^5+2q^6\right.\right. \right. \nonumber\\
&~&\left. \left.\left. +q^7+2q^8+q^{10}\right)
-\lambda^3q^{\frac{9}{2}}
\left(2+8q+15q^2+32q^3+45q^4+65q^5 \right. \right. \right. \nonumber\\
&~& \left. \left. \left.+75q^6 +85q^7 + 82q^8+77q^9
+65q^{10}+51q^{11}+37q^{12}+25q^{13}  \right. \right. \right. \nonumber\\
&~&\left. \left. \left.+ 14q^{14}+7q^{15}+3q^{16}\right) +q^{\frac{11}{2}}
\left(2+5q^2+3q^3 +7q^4+5q^5 +10q^6  \right. \right. \right. \nonumber\\
&~&\left. \left. \left. +5q^7+8q^8+ 5q^9+5q^{10} 
+2q^{11}+3q^{12}+q^{14}\right)
-\lambda q^{\frac{9}{2}}
\left(2+4q \right. \right. \right. \nonumber\\
&~&\left. \left. \left. +9q^2+15q^3+23q^4+27q^5+36q^6+37q^7 
+ 36q^8 \right. \right. \right. \nonumber\\
&~&\left. \left. \left. +34q^9+29q^{10}
+21q^{11}+16q^{12}+10q^{13}+5q^{14}+3q^{15}+q^{16}\right) \right. \right.
\nonumber\\
&~& \left. \left. -\lambda^5q^{\frac{13}{2}}
\left(2+3q+8q^2+10q^3 +15q^4+14q^5+16q^6+15q^7 \right. \right. \right.
\nonumber\\
&~&\left. \left. \left.+12q^8+10q^9+9q^{10}+5q^{11}+3q^{12}+2q^{13}\right)
+\lambda^4q^{\frac{11}{2}}
\left(4+6q+18q^2 \right. \right. \right. \nonumber\\
&~&\left. \left. \left. +24q^3+39q^4+43q^5+53q^6+46q^7+ 
50q^8+37q^9 +33q^{10}+22q^{11} \right. \right. \right. \nonumber\\
&~&\left. \left. \left. +18q^{12}+7q^{13}+ 
6q^{14}+q^{15}\right)+\lambda^2q^{\frac{9}{2}}
\left(4+6q+20q^2+27q^3\right.\right.\right.\nonumber\\&~&\left.\left.\left.
+48q^4+55q^5+76q^6+72q^7+80q^8+67q^9+62q^{10}+43q^{11}+36q^{12}
\right.\right.\right.\nonumber\\&~&\left.\left.\left.
+19q^{13}+14q^{14}+5q^{15}+3q^{16}\right)\right\}\right]
\end{eqnarray}
And, for $R={\twover}$, we obtain
\begin{eqnarray}
f_{\twover}(3_1\#5_1)&=&\denomneg
\left[\lambda^5q^{-\frac{23}{2}}{\left(\lambda-1\right)}^2
\left(\lambda-q\right)\left(\lambda q-1\right) 
\left\{1+3q^2+2q^3+\right.\right.\nonumber\\
&~&\left.\left. 5q^4 + 5q^5 + 8q^6 + 5q^7 + 
10q^8+5q^9+7q^{10}+3q^{11}+5q^{12}+2q^{14} 
\right.\right.\nonumber\\&~&\left.\left.
-\lambda\left(1+q+2q^2+4q^3+4q^4+6q^5+8q^6+8q^7
+7q^8+9q^9
\right.\right.\right.\nonumber\\&~&\left.\left.\left.
+5q^{10}+5q^{11}+2q^{12}+2q^{13}\right)+
\lambda^2\left(q+2q^3q^4+2q^5+q^6
+3q^7\right.\right.\right.\nonumber\\&~&\left.\left.\left.
+q^8+q^9+q^{10}+q^{11}\right)\right\}\right] 
\end{eqnarray}
\subsection{Invariants for the Links}   
In this subsection, we will consider invariants obtained for two 
and three component links carrying defining representations.
We have seen that $f_{\one}$ for knots are exactly the
knot polynomials in the defining representation (\ref {onebox}). 
However, naively extending this equality to links does not match
with Ooguri-Vafa conjecture. 

For the $2$-component link denoted by $2_1^2$ 
(also called the Hopf link in the knot theory literature) we have
from eq. (\ref{3.41}) the following link invariant in defining representation:
\begin{equation}
V_{\one,\one}(2_1^2)=\frac{\lambda^{1\over 2}-\lambda^{-{1\over 2}}}
{\left(q^{1\over 2}-q^{-{1\over 2}}\right)^2}
\left[q^{-1}\lambda^{1\over 2}\left\{-1+q\left(1-q+\lambda\right)
\right\}\right]
\label{pa1}
\end{equation}
Similarly, from the methods presented in the last section, we find
that for the $2$-component links $4_1^2$ (\ref {3.42}) and 
$6_1^2$ (\ref {3.43}), the link invariants in defining 
representations are respectively,
\begin{eqnarray}
V_{\one,\one}(4_1^2)&=&
\frac{\lambda^{1\over 2}-\lambda^{-{1\over 2}}}
{\left(q^{1\over 2}-q^{-{1\over 2}}\right)^2}
\left[q^{-2}\lambda^{3\over 2}
\left(-1+q-q^2+q^3-q^4+\lambda q-\lambda q^2+\lambda q^3 
\right)\right] 
\label{pa2}\\
V_{\one,\one}(6_1^2)&=&
\frac{\lambda^{1\over 2}-\lambda^{-{1\over 2}}}
{\left(q^{1\over 2}-q^{-{1\over 2}}\right)^2}
\left[q^{-3}\lambda^{5\over 2}
\left\{-1+q-q^2+q^3-q^4+q^5-q^6+\right.\right.\nonumber\\
&~&\left.\left.\lambda\left(q-q^2+ 
q^3-q^4+q^5\right)\right\}\right] 
\label{pa3}
\end{eqnarray}
Similarly, for three component links $6_1^3, 6_3^3$, 
the invariant in defining representation \cite {lick}
\footnote{We make an appropriate change of variable here. The 
variables $l$ and $m$, defined in \cite{lick} are related to
$q$ and $\lambda$ as $l=i\lambda^{1\over 2};~m=i\left(
q^{1\over 2}-q^{-{1\over 2}}\right)$}
\begin{eqnarray}
V_{\one,\one,\one}(6_1^3)&=&
\frac{\lambda^{1\over 2}-\lambda^{-{1\over 2}}}
{\left(q^{1\over 2}-q^{-{1\over 2}}\right)^3}
\left(q^{-3}\lambda^{-2}\right)
\left[q+q^2\left(q-1\right)
\left(3-2q+q^2\right) \right. \label {pa4}\\ 
&~& \left. +\lambda^2q\left(1-q+q^2\right)^2  
-\lambda\left\{1+\left(q^2-1\right)
\left(3+q-2q^3+q^4\right) \right\} \right] \nonumber\\
V_{\one,\one,\one}(6_3^3)&=&
\frac{\lambda^{1\over 2}-\lambda^{-{1\over 2}}}
{\left(q^{1\over 2}-q^{-{1\over 2}}\right)^3}
\left(q^{-2}\lambda^2\right)
\left[q\left(1-q+q^2\right)- 
\lambda^{-1}\left(1-q+^2\right)
\left(1+q^2\right)\right.\nonumber\\&~&\left.
+\lambda^{-2}q\left(2-3q+2q^2\right)\right]
\label{pa5}
\end{eqnarray}
From these link invariants in 
defining representation, it is clear that they cannot
be directly equated to new polynomial invariant 
$f_{\one, \one, \ldots}$ as there are extra powers
of $q^{1/2} - q^{-1/2}$ in the denominator, violating the
Ooguri-Vafa conjecture. 
Recently Labastida et al \cite {lablatest} (see note added)
have derived new polynomial invariants for links involving
link invariants in Chern-Simons theory. Hence using their
results and the link invariants in defining representations,
$f_{\one, \one, \ldots}$ can be shown to obey the 
conjecture.

\section{Discussion and Conclusions}
In this paper, we have performed non-trivial tests of the conjecture
in \cite{gv2,ov} relating topological string amplitudes on the
resolved conifold geometry to Chern-Simons gauge theory on $S^3$. 
The elegant group theoretic procedure developed in 
Ref. \cite{lm} to determine topological string amplitudes
from vevs of Wilson loops of Chern-Simons theory enables us
to verify the conjecture for a number of knots upto nine 
crossings.

The methods developed in \cite{gkr,thesi}
established a simple and powerful way of computing
generalised knot and link invariants in $SU(N)$ Chern-Simons theory. 
In particular, we have gone much beyond
the results presented in \cite{lm} using our tools, and have found
beautiful agreement with the conjecture of \cite{ov} for new polynomial
invariants for these knots. We believe that our results firmly
establishes the said conjecture. Further, this is the first time 
the integer coefficients of these (new) polynomial
invariants in knot theory  has meaning in terms of 
certain D-brane computations in closed topological string theory.

We have the machinery to obtain knot and link invariants 
in Chern-Simons theory. However, we have been able 
relate only knot invariants to closed topological string amplitudes.
It would be extremely interesting to generalise the topological string theory
computation to objects that are $n$-component links in the dual theory,
and to check the new polynomial invariants that arise in this case using
the methods developed in this paper. 
 
Another interesting direction will be to explore the relationship 
of perturbative Chern-Simons invariants - namely, Vassiliev invariants to
objects in the dual string side. We leave the study of such issues to
a future publication.

\vskip.5cm
\noindent
{\bf Acknowledgments}: P.R would like to thank CSIR for the grant.
\vskip.5cm
\noindent
{\bf Note Added}: After this work appeared in the bulletin board,
the very important paper \cite{lablatest} by Labastida, Marino 
and Vafa appeared, which has treated the case of multi-component 
links observables in large N Chern-Simons theory and dual topological 
string picture, in great details. 

\newpage
\section{Appendix}
In this appendix, we present the explicit results for the
values of $f_R$ for $R$ containing three boxes in its Young
tableau. We will do so for the $7_1$ knot, i.e
a two stranded braid with seven crossings. The results
are
\begin{eqnarray}
f_{\threehor}(7_1)&=&\denomneg
\left[q^{-7}\lambda^{1\over 2}
\left(1-q+q^2\right)
\left(1+q^4\right)\right.\nonumber\\&~&\left.
\left(1+q+q^2+q^3+q^4+q^5+q^6\right) 
{\left(\lambda-1\right)}^2\left(\lambda-q\right)
\left(\lambda q-1\right)\right.\nonumber\\&~&\left.
\left\{q^2\lambda^2\left(1+q^2\right)
\left(1+q^3+q^6\right)
\left(1+2q+q^2+q^6+q^{10}\right)\right.\right.
\nonumber\\ &~&\left.\left.
-\lambda\left(
1+2q+2q^2+3q^3+4q^4+5q^5+7q^6+8q^7+7q^8+ 
6q^9+5q^{10}
\right.\right.\right.\nonumber\\&~&\left.\left.\left.
+4q^{11}
+4q^{12}+4q^{13}+2q^{14} + 
2q^{15}+2q^{16}+q^{17}+q^{18}+q^{19}\right)
\right.\right.\nonumber\\ &~&\left.\left.
+q^2\left(1+2q+2q^2+3q^3+3q^4+2q^5+3q^6+3q^7+2q^8+ 
3q^9+q^{10}
\right.\right.\right.\nonumber\\&~&\left.\left.\left.
+q^{11}
+q^{12}+q^{13}+q^{15}\right)\right\}\right] \\
f_{\threever}(7_1)&=&\denomneg
\left[q^{-\frac{63}{2}}\lambda^{15\over 2}\left(1-q+q^2\right)
\left(1-q^{1\over 2}+q-q^{\frac{3}{2}}+q^2
-q^{\frac{5}{2}}+q^3\right)\right.\nonumber\\&~&\left. 
\left(1+q^{1\over 2}+q+q^{\frac{3}{2}}+q^2 + 
q^{\frac{5}{2}}+q^3 \right) \left(1+q^4\right) 
\left\{q^{\frac{17}{2}}+q^{\frac{21}{2}} + q^{\frac{23}{2}} + 
q^{\frac{25}{2}}+q^{\frac{27}{2}}
\right.\right.
\nonumber\\&~&\left.\left. 
+3q^{\frac{29}{2}} 
+2q^{\frac{31}{2}}+3q^{\frac{33}{2}}+3q^{\frac{35}{2}} + 
2q^{\frac{37}{2}}+3q^{\frac{39}{2}}+3q^{\frac{41}{2}} + 
2q^{\frac{43}{2}}+2q^{\frac{45}{2}}+q^{\frac{47}{2}}
\right.\right.\nonumber\\&~&\left.\left.
+\lambda\left(-q^{\frac{13}{2}}-2q^{\frac{15}{2}}
-3q^{\frac{17}{2}}-4q^{\frac{19}{2}}- 
5q^{\frac{21}{2}}-6q^{\frac{23}{2}}- 
8q^{\frac{25}{2}}-10q^{\frac{27}{2}}- 
13q^{\frac{29}{2}}
\right.\right.\right.
\nonumber\\&~&\left.\left.\left.
-15q^{\frac{31}{2}} 
-17q^{\frac{33}{2}}-18q^{\frac{35}{2}} - 
18q^{\frac{37}{2}}-18q^{\frac{39}{2}} - 
16q^{\frac{41}{2}} 
\right.\right.\right.\nonumber\\&~&\left.\left.\left.
-13q^{\frac{43}{2}} - 
10q^{\frac{45}{2}}-6q^{\frac{47}{2}} 
-3q^{\frac{49}{2}}-q^{\frac{51}{2}}\right) 
\right.\right.\nonumber\\&~&\left.\left.
+\lambda^2\left(2q^{\frac{11}{2}}+3q^{\frac{13}{2}}+ 
7q^{\frac{15}{2}}+8q^{\frac{17}{2}}+ 
12q^{\frac{19}{2}}+ 13q^{\frac{21}{2}}+ 
\right.\right.\right.\nonumber\\&~&\left.\left.\left.
18q^{\frac{23}{2}}+21q^{\frac{25}{2}}+ 
28q^{\frac{27}{2}}+32q^{\frac{29}{2}}+ 
39q^{\frac{31}{2}}+43q^{\frac{33}{2}}+ 
48q^{\frac{35}{2}}+ 
\right.\right.\right.\nonumber\\&~&\left.\left.\left.
48q^{\frac{37}{2}}+ 
46q^{\frac{39}{2}}+40q^{\frac{41}{2}}+ 
32q^{\frac{43}{2}}+24q^{\frac{45}{2}}+ 
16q^{\frac{47}{2}}+9q^{\frac{49}{2}}
\right.\right.\right.\nonumber\\&~&\left.\left.\left.
+4q^{\frac{51}{2}} + q^{\frac{53}{2}}\right)
+\lambda^3\left(-q^{\frac{9}{2}}-4q^{\frac{11}{2}} - 
6q^{\frac{13}{2}}-10q^{\frac{15}{2}}- 
14q^{\frac{17}{2}}-16q^{\frac{19}{2}}- 
\right.\right.\right.\nonumber\\&~&\left.\left.\left.
20q^{\frac{21}{2}}-26q^{\frac{23}{2}}- 
30q^{\frac{25}{2}}-38q^{\frac{27}{2}}- 
46q^{\frac{29}{2}}-52q^{\frac{31}{2}}- 
60q^{\frac{33}{2}} 
\right.\right.\right.\nonumber\\&~&\left.\left.\left.
-66q^{\frac{35}{2}}- 
65q^{\frac{37}{2}}-62q^{\frac{39}{2}}- 
54q^{\frac{41}{2}}-42q^{\frac{43}{2}}- 
32q^{\frac{45}{2}}-22q^{\frac{47}{2}}- 
\right.\right.\right.\nonumber\\&~&\left.\left.\left.
12q^{\frac{49}{2}}-6q^{\frac{51}{2}} - 
2q^{\frac{53}{2}}\right)+\lambda^4\left(2q^{\frac{9}{2}} + 
3q^{\frac{11}{2}}+7q^{\frac{13}{2}}+ 
8q^{\frac{15}{2}}+12q^{\frac{17}{2}}+ 
\right.\right.\right.\nonumber\\&~&\left.\left.\left.
13q^{\frac{19}{2}}+17q^{\frac{21}{2}}+ 
19q^{\frac{23}{2}}+25q^{\frac{25}{2}}+ 
29q^{\frac{27}{2}}+36q^{\frac{29}{2}}+ 
\right.\right.\right.\nonumber\\&~&\left.\left.\left.
40q^{\frac{31}{2}}+47q^{\frac{33}{2}}+ 
49q^{\frac{35}{2}}+50q^{\frac{37}{2}}+ 
46q^{\frac{39}{2}}+40q^{\frac{41}{2}}+ 
\right.\right.\right.\nonumber\\&~&\left.\left.\left.
32q^{\frac{43}{2}}+24q^{\frac{45}{2}}+ 
16q^{\frac{47}{2}}+9q^{\frac{49}{2}}+ 
4q^{\frac{51}{2}}+q^{\frac{53}{2}}\right)
\right.\right.\nonumber\\&~&\left.\left.
+\lambda^5\left(-q^{\frac{9}{2}}-2q^{\frac{11}{2}} - 
3q^{\frac{13}{2}}-4q^{\frac{15}{2}}- 
5q^{\frac{17}{2}}-6q^{\frac{19}{2}}- 
7q^{\frac{21}{2}}-8q^{\frac{23}{2}}- 
10q^{\frac{25}{2}}
\right.\right.\right.\nonumber\\&~&\left.\left.\left.
-12q^{\frac{27}{2}}- 
15q^{\frac{29}{2}}-17q^{\frac{31}{2}}- 
19q^{\frac{33}{2}}-20q^{\frac{35}{2}}- 
19q^{\frac{37}{2}}-18q^{\frac{39}{2}}- 
16q^{\frac{41}{2}} 
\right.\right.\right.\nonumber\\&~&\left.\left.\left.
-13q^{\frac{43}{2}}- 
10q^{\frac{45}{2}}-6q^{\frac{47}{2}} - 
3q^{\frac{49}{2}}-q^{\frac{51}{2}}\right)  
\right.\right.\nonumber\\&~&\left.\left.
+\lambda^6\left(q^{\frac{11}{2}}+q^{\frac{15}{2}}+ 
q^{\frac{17}{2}}+q^{\frac{19}{2}}+ 
q^{\frac{21}{2}}+2q^{\frac{23}{2}}+ 
q^{\frac{25}{2}}+2q^{\frac{27}{2}}+ 
3q^{\frac{29}{2}}+3q^{\frac{31}{2}}+ 
\right.\right.\right.\nonumber\\&~&\left.\left.\left.
3q^{\frac{33}{2}}+4q^{\frac{35}{2}}+ 
2q^{\frac{37}{2}}+3q^{\frac{39}{2}}+ 
3q^{\frac{41}{2}}+2q^{\frac{43}{2}}+ 
\right.\right.\right.\nonumber\\&~&\left.\left.\left.
2q^{\frac{45}{2}}+q^{\frac{47}{2}}\right)\right\}
\right] \\
f_{\mixed}(7_1)&=&\denom
\left[q^{-\frac{61}{2}}\lambda^{15\over 2}\left(1-q+q^2\right) 
\left(1-q^{1\over 2}+q-q^{\frac{3}{2}}+q^2-q^{\frac{5}{2}}+ 
q^3\right)
\left(1+q^{1\over 2} \right. \right. \nonumber\\
&~&\left. \left. +q+q^{\frac{3}{2}}+q^2 + 
q^{\frac{5}{2}}+q^3\right)\left(1+q^4\right) 
\left\{q^{\frac{39}{2}}+2q^{\frac{41}{2}}+3q^{\frac{43}{2}} + 
3q^{\frac{45}{2}}+4q^{\frac{47}{2}} 
\right.\right.\nonumber\\&~&\left.\left.
+3q^{\frac{49}{2}}+ 
4q^{\frac{51}{2}}+3q^{\frac{53}{2}}+3q^{\frac{55}{2}}+ 
2q^{\frac{57}{2}}+q^{\frac{59}{2}}+\lambda\left(-q^{\frac{35}{2}}
-3q^{\frac{37}{2}}-7q^{\frac{39}{2}}- 
12q^{\frac{41}{2}} 
\right.\right.\right.\nonumber\\&~&\left.\left.\left.
-16q^{\frac{43}{2}}-
20q^{\frac{45}{2}}-23q^{\frac{47}{2}}- 
23q^{\frac{49}{2}}-23q^{\frac{51}{2}}- 
20q^{\frac{53}{2}}-16q^{\frac{55}{2}}- 
12q^{\frac{57}{2}}- 
\right.\right.\right.\nonumber\\&~&\left.\left.\left.
7q^{\frac{59}{2}}-
3q^{\frac{61}{2}}- q^{\frac{63}{2}}\right) 
+\lambda^2\left(q^{\frac{33}{2}}+4q^{\frac{35}{2}}+ 
11q^{\frac{37}{2}}+19q^{\frac{39}{2}} + 
31q^{\frac{41}{2}}+
\right.\right.\right.\nonumber\\&~&\left.\left.\left.
40q^{\frac{43}{2}}+
52q^{\frac{45}{2}}+58q^{\frac{47}{2}}+ 
62q^{\frac{49}{2}}+58q^{\frac{51}{2}}+ 
52q^{\frac{53}{2}}+ 
\right.\right.\right.\nonumber\\&~&\left.\left.\left.
40q^{\frac{55}{2}}+31q^{\frac{57}{2}} +
19q^{\frac{59}{2}}+
11q^{\frac{61}{2}}+ 4q^{\frac{63}{2}} + 
q^{\frac{65}{2}}\right)
\right.\right.\nonumber\\&~&\left.\left.
+\lambda^3\left(-2q^{\frac{33}{2}}- 
7q^{\frac{35}{2}}-16q^{\frac{37}{2}}- 
28q^{\frac{39}{2}}-42q^{\frac{41}{2}}-
56q^{\frac{43}{2}}-70q^{\frac{45}{2}} - 
\right.\right.\right.\nonumber\\&~&\left.\left.\left.
80q^{\frac{47}{2}}-84q^{\frac{49}{2}}- 
80q^{\frac{51}{2}}-70q^{\frac{53}{2}}- 
56q^{\frac{55}{2}}-
\right.\right.\right.\nonumber\\&~&\left.\left.\left.
42q^{\frac{57}{2}}-
28q^{\frac{59}{2}}-16q^{\frac{61}{2}}- 
7q^{\frac{63}{2}}-2q^{\frac{65}{2}}\right) 
\right.\right.\nonumber\\&~&\left.\left.
+\lambda^4\left(q^{\frac{33}{2}}+6q^{\frac{35}{2}}+ 
12q^{\frac{37}{2}}+23q^{\frac{39}{2}}+
32q^{\frac{41}{2}}+44q^{\frac{43}{2}}+ 
53q^{\frac{45}{2}}+62q^{\frac{47}{2}}+ 
\right.\right.\right.\nonumber\\&~&\left.\left.\left.
63q^{\frac{49}{2}}+62q^{\frac{51}{2}}+
53q^{\frac{53}{2}}+44q^{\frac{55}{2}}+
32q^{\frac{57}{2}}+23q^{\frac{59}{2}} + 
12q^{\frac{61}{2}}+6q^{\frac{63}{2}} + 
q^{\frac{65}{2}}\right) 
\right.\right.\nonumber\\&~&\left.\left.
+\lambda^5\left(-2q^{\frac{35}{2}}- 
5q^{\frac{37}{2}}-9q^{\frac{39}{2}}-
14q^{\frac{41}{2}}-18q^{\frac{43}{2}}- 
22q^{\frac{45}{2}}-25q^{\frac{47}{2}}- 
\right.\right.\right.\nonumber\\&~&\left.\left.\left.
25q^{\frac{49}{2}}-25q^{\frac{51}{2}}- 
22q^{\frac{53}{2}}-18q^{\frac{55}{2}}-
14q^{\frac{57}{2}}-9q^{\frac{59}{2}}- 
\right.\right.\right.\nonumber\\&~&\left.\left.\left.
5q^{\frac{61}{2}}-2q^{\frac{63}{2}}\right)
+\lambda^6\left(q^{\frac{37}{2}}+q^{\frac{39}{2}}+ 
3q^{\frac{41}{2}}+
\right.\right.\right.\nonumber\\&~&\left.\left.\left.
3q^{\frac{43}{2}}+
4q^{\frac{45}{2}}+4q^{\frac{47}{2}}+ 
4q^{\frac{49}{2}}+4q^{\frac{51}{2}}+ 
4q^{\frac{53}{2}}+
\right.\right.\right.\nonumber\\&~&\left.\left.\left.
3q^{\frac{55}{2}}+
3q^{\frac{57}{2}}+ q^{\frac{59}{2}}+ 
q^{\frac{61}{2}}\right)\right\}\right] 
\end{eqnarray}
 
\end{document}